\documentclass{aastex631}

\shorttitle{High Energy Emission of TXS0506+056}
\shortauthors{Wang et al.}

\graphicspath{{./}{figures/}}
\usepackage{appendix}
\usepackage{amsmath}
\usepackage{color}
\usepackage{multirow}
\begin{document}

\title{A Unified Model for Multi-epoch Neutrino Events and Broadband Spectral Energy Distribution of TXS~0506+056}

\author{Zhen-Jie Wang}
\affiliation{Department of Astronomy, Xiamen University, Xiamen, Fujian 361005, China}

\author[0000-0003-1576-0961]{Ruo-Yu Liu}
\affiliation{School of Astronomy and Space Science, Xianlin Road 163, Nanjing University, Nanjing 210023, China}
\affiliation{Key laboratory of Modern Astronomy and Astrophysics (Nanjing University), Ministry of Education, Nanjing 210023, China}

\author{Ze-Rui Wang}
\affiliation{College of Physics and Electronic Engineering, Qilu Normal University, Jinan 250200, China}

\author[0000-0003-4874-0369]{Junfeng Wang}
\affiliation{Department of Astronomy, Xiamen University, Xiamen, Fujian 361005, China}

\correspondingauthor{Ruo-yu Liu; Junfeng Wang}
\email{ryliu@nju.edu.cn; jfwang@xmu.edu.cn}
\begin{abstract}

The blazar TXS~0506+056 has been proposed as a high-energy neutrino emitter. However, it has been shown that the standard one-zone model cannot produce sufficiently high neutrino flux due to constraints from the X-ray data, implying more complex properties of the radiation zones in the blazar than that described by the standard one-zone model. In this work we investigate multi-epoch high-energy muon neutrino events associated with the blazar TXS~0506+056 occured in 2014--2015, 2017--2018, 2021--2022 and 2022--2023, respectively. We applied the so-called ``stochastic dissipation model'' to account for the neutrino-blazar associations detected in the four epochs simultaenously. This model describes a scenario in which the emission of the blazar arise from the superimposition of two components: a persistent component related to the quasi-stable state of the blazar and a transient component responsible for the sudden enhancement of the blazar's flux, either in electromagnetic radiation or in neutrino emission. The latter component could form at a random distance along the jet by a strong energy dissipation event. Under such assumption, the multi-epoch broadband spectral energy distribution (SED) can be well explained and the expected number of high-energy neutrino events is statistically realistic. The expected number of neutrino events in half-year is around 8.2, 0.07, 0.73 and 0.41, corresponding to the epoch in 2014--2015, 2017--2018, 2021--2022 and 2022--2023, respectively. Hence, our model self-consistently explains the episodic neutrino emission from TXS~0506+056.

\end{abstract}


\section{Introduction} \label{sec:intro}

    The origin of high-energy neutrinos is still unclear \citep{Aartsen_2013, IceCube_2013}. As the most powerful persistent emitters of electromagnetic (EM) radiation in the universe, blazars have long been suggested as potential accelerators of high-energy cosmic-ray protons (or nuclei) and sources of high-energy neutrinos \citep[e.g.,][]{1993A&A...269...67M, 1995APh.....3..295M, 1996SSRv...75..341S, Halzen_2002, 2001PhRvL..87v1102A, Murase_2014PhRvD..90b3007M}. The discovery of a possible correlation between high-energy neutrino events and blazars supports this speculation. \citep{2018Sci...361.1378I, 2018Sci...361..147I}. Many researchers have studied these events in the framework of the so-called photohadronic interaction model \citep{Ansoldi_2018, 2018ApJ...864...84K, 2018ApJ...865..124M, 2018MNRAS.480..192P, Padovani_2019MNRA,  Cerruti_2019, 2019NatAs...3...88G, 2019ApJ...881...46R, Rodrigues_2019ApJ, Xue_2019, IceCube:2021oqh} and hadronuclear interaction model \citep{2018ApJ...866..109S, Liu_2019, 2019PhRvD..99j3006B}. In these hadronic processes, the produced secondary EM particles can induce EM cascades, contributing to EM radiation from the optical band to the X-ray band or even to the MeV gamma-ray band \citep{Xue_2021, 2019ApJ...881...46R}. On the other hand, the low-energy radiation of a blazar can be the target radiation field for the photohadronic interactions \citep{2008PhRvD..78c4013K, Murase_2014PhRvD..90b3007M, Murase_2022arXiv220203381M}. In this sense, the neutrino emission from a blazar is related to its broadband radiation.  

    The first blazar associating with neutrino events at a significance above $3\sigma$, TXS 0506+056, is a known gamma-ray blazar at redshift $z=0.3365$ \citep{Paiano_2018}. On 2017 September 22, a 290 TeV muon neutrino event (known as IC-170922A), was detected by IceCube \citep{ 2018Sci...361.1378I} from the direction of the blazar, when the blazar underwent a GeV gamma-ray flare. The chance coincidence of the association is rejected at a significance level of approximately $3\sigma$. IceCube also found indication of a neutrino outburst with $13\pm 5$ muon neutrino event in 32\,TeV -- 3.6\,PeV from this blazar over $\sim 160$ days from 2014–2015 at a significance of $3.5\sigma$ level when examining over nine years of archival data \citep{2018Sci...361..147I}. Great efforts have been made to understand the theoretical correlation between the neutrino emission and the blazars \citep{1995APh.....3..295M, 1996SSRv...75..341S, 2001PhRvL..87v1102A, Murase_2014PhRvD..90b3007M, Murase_2022arXiv220203381M}. An consensus has been reached that the standard one-zone model cannot produce sufficiently high neutrino flux due to constraints from the X-ray data \citep{2018ApJ...864...84K, 2019NatAs...3...88G, Rodrigues_2019ApJ}, and the detection of IC-170922A has to be interpreted as the Eddington bias \citep{2019A&A...622L...9S} or an upward fluctuation in the detection, given an anticipated event rate of 0.01/year. The problem for the 2014-2015 outburst is more serious. \citet{Rodrigues_2019ApJ} found that the standard one-zone model can at most generate 4.9 events in half a year. This discrepancy cannot be ascribed to the Eddington bias because of the large difference in the event number between expectation and observation. More complex models invoking two or even more radiation zones are needed if we want to reproduce a more comfortable neutrino event rate to account for these association, especially for the 2014-2015 outburst \citep{Xue_2021, Rodrigues_2019ApJ}.
    Later, an independent search for point-like sources in the northern hemisphere using ten years of IceCube data revealed that TXS~0506+056 is coincident with the second hottest-spot of neutrino event excess \citep{IceCube_2020}. In addition, interestingly, a cascade neutrino event with the estimated energy of $224\pm75$\,TeV, GVD-210418CA, from the direction of TXS 0506+056 was detected by the new Baikal-GVD neutrino telescope in April 2021 \citep{Baikal-GVD_2021}.  On September 18, 2022, a $\sim$ 170 TeV neutrino, IC-220918A, was reported in spatial coincident with TXS 0506+056 again by IceCube \citep{IceCube_2022}. Although the latter two neutrino events have not been studied in detail, the maximum expected neutrino event rate under the standard one-zone model would be similar to that for IC-170922A, i.e., $\sim$0.01/year, because the X-ray fluxes in the latter two epochs are comparable to (or slightly lower than) that in the epoch IC-170922A. If the latter two neutrino events are truly associated with TXS 0506+056, the standard one-zone model can be ruled out, because the joint possibility would be unreasonably low if all these associations are explained as lucky detections. 

    The motivation of this work is to explore whether the multi-messenger emission can be self-consistently accounted for in all four epochs. Also, as can be seen later, the SED in these four epochs are different. Hence, it is also interesting to investigate if there exists an inherent relation among the four epochs with neutrino detection. \citet{Wang_Ze-Rui_2022} proposed the so-called ``stochastic dissipation model'' to explain the orphan and multiwavelength flares from blazars in a unified framework. They speculate that a blazar's emission is composed of contribution from numerous radiation zones (or blobs), formed by certain energy dissipation events, along the jet. Therefore, the blazar's emission consists of two components during a flare: (i) a quasi-stable component that arises from the superposition of numerous but comparatively weak dissipation zones along the jet, forming a persistent emission of the blazar which is statistically identical among different epochs \citep{Meyer_2019}; (ii) a transient component, which is responsible for the sudden enhancement of the blazar flux, generated at a random distance along the jet by a strong energy dissipation event \citep{Giannios_2006A&A, Nalewajko_10.1111, Begelman_1998ApJ, McKinney_2009MNRA}. In this work, we explore the broadband SED and the neutrino emission of TXS 0506+056 in multi-epochs in the framework of this model, assuming the presence of high-energy protons in the dissipation events. 
    
    The rest of the paper is organized as follows. The observation and data reduction process are described in Section 2. The model and method for calculating the radiation is described in Section 3. In Section 4 we show the SED modeling and results. Finally we present discussion and summarize our findings in Section 5. Throughout the paper, $H_0=$71 km s$^{-1}$ Mpc$^{-1}$, $\Omega_{m}=0.27$, and $\Omega_{\Lambda}=0.73$ are adopted.

\section{OBSERVATIONS AND DATA REDUCTION}

    For the 2014–2015 neutrino outburst and the IC-170922A flare, the multiwavelength data are from \citet{Rodrigues_2019ApJ} and \citet{2018Sci...361.1378I}, respectively. For the periods covering the 2021-2022 neutrino event GVD-210418CA and the 2022-2023 neutrino event IC-220918A, the multiwavelngth data are processed by ourselves in this work. 

\subsection{$\gamma$-Ray Observations by Fermi-LAT}

    We first obtained the Pass 8 data from the Fermi Science Support Center (FSSC). The data are selected within a 15$^{\circ}$ region of interest (ROI) centered on the radio position of TXS 0506+056 (R.A. = 77.3582$^{\circ}$, Decl. = 5.69315$^{\circ}$; \cite{Beasley_2002ApJS}). For the 2021-2022 neutrino events, the temporal coverage of the data is from 2021 January 01 to 2021 December 31 (MJD 59215–59580) of about 1.0 yr. For the 2022-2023 neutrino events, the temporal coverage of the data is from 2022 January 01 to 2022 December 31 (MJD 59580-59944) of about 1.0 yr. A binned likelihood analysis is performed for the $\gamma$-rays of 4FGL J0509.4+0542 using the publicly available software {\tt Fermitools} v.2.0.8. In our analysis only the $\gamma$-ray photons in the energy range of 0.1-300 GeV and satisfying the standard data quality selection criteria “$(\rm DATA\underline{~}QUAL \textgreater 0)\&\&(LAT\underline{~}CONFIG == 1)$” are considered. A zenith angle cut of 90$^{\circ}$ is set to avoid the $\gamma$-ray contamination causing by the Earth limb. We bin the data with a pixel size of 0.2$^{\circ}$ in space and 25 logarithmical energy bins. The background models include all $\gamma$-ray sources listed in the 4FGL-DR3 Catalog and the Galactic diﬀuse component (gll$\underline{~}$iem\underline{~}v07.fits) as well as the isotropic emission (iso$\underline{~}$P8R3$\underline{~}$SOURCE$\underline{~}$V3$\underline{~}$v1.txt). The $\rm P8R3\underline{~}SOURCE\underline{~}V3$ set of instrument response functions (IRFs) is used. 

    The model for fitting the spectrum of 4FGL J0509.4+0542 in our analysis is a single power-law function, 
    \begin{equation}
    \frac{dN}{dE} = N_{0}\left(\frac{E}{E_{b}}\right)^{-\Gamma_{\gamma}},
    \end{equation}
    where $\Gamma_{\gamma}$ is the photon spectral index, and $E_{b}$ is the scale parameter of photon energy \citep{Massaro_2004A&A}. The spectral parameters of all sources lying within 8$^{\circ}$ are left free, whereas the parameters of those sources lying beyond 8$^{\circ}$ are fixed to their 4FGL-DR3 values. Also, the normalization parameters of the standard Galactic and isotropic background templates are set free in the likelihood fit. The significance of the $\gamma$-ray detection is quantified by adopting the maximum likelihood test statistic (TS), TS = 2($\mathcal{L}_{1}$-$\mathcal{L}_{0}$), where $\mathcal{L}_{1}$ and $\mathcal{L}_{0}$ are the maximum likelihood values for the models with and without the target source, respectively. The sources with $\rm TS < 16$ are removed from further analysis. The results of the analysis and SED fitting are shown in Table 1 and Figure 7-8. The TS maps are shown in Figure 11-12.

\subsection{X-Ray Observations with Swift-XRT}

    The X-ray telescope (XRT) on board the Swift satellite observed the source in the photon counting mode with exposure times of 21.1 ks and 2.5 ks, in 2021-2022 and 2022-2023 respectively. We perform standard filtering and data analysis using HEASOFT (v 6.29). To extract the source spectrum, we select a circular region of 30$^{\prime\prime}$, centered at the target. The background is estimated from an annular cell. The source spectrum is binned to have at least 20 counts per bin. The SED fitting process is performed in XSPEC and the $\chi^{2}$ minimization technique is adopted for spectral analysis. The associated errors are calculated at the $90\%$ confidence level. The spectrum is fitted by a single power law absorbed by two absorption components, one is absorption at z = 0 with the neutral hydrogen column density fixed at Galactic value $N_{gal}^{H}=1.55\times10^{21}\rm cm^{-2}$\citep{Willingale_2013MNRAS}, the other is an extragalactic foreground absorption $N_{int}^{H}$ at a redshift of z = 0.3365 with column density set free. An extragalactic foreground absorption column density of $N_{int}^{H}$ would be obtained by SED fitting. The complete process of building Swift-XRT products are described in \url{https://www.swift.ac.uk/user_objects/}. The results of the analysis and SED fitting are shown in Table 2 and Figure 9-10.

\subsection{Optical Observations with ALeRCE}

    The Automatic Learning for the Rapid Classification of Events \citep[ALeRCE,][]{Forster_2021} broker light curve classifier \citep[LCC,][]{Sanchez-Saez_2021} is processing the alert stream from the Zwicky Transient Facility (ZTF) and which aims to become a Community Broker for the Vera C. Rubin Observatory and its Legacy Survey of Space and Time (LSST), as well as other large etendue survey telescopes. The goal of the LCC is to provide a fast classification of transient and variable objects by applying a balanced random forest algorithm. The LCC produced ariability features and colors obtained from AllWISE and ZTF photometry in g and r band. The complete set of features are described in \url{http://alerce.science/features/}. In this work, we extracted variability features from the g and r bands to follow the neutrino events of TXS 0506+056 during the periods of 2021-2022 and 2022-2023, respectively.

    \begin{deluxetable*}{cccc}
    \tablenum{1}
    \tablecaption{Results of the Analysis of Fermi-LAT Observations of TXS~0506+056\label{tab:messier}}
    \tablewidth{0pt}
    \tablehead{
        \colhead{year} & \colhead{$\rm \Gamma_{\gamma}$} & \colhead{$F_{\gamma}$} & \colhead{$\rm TS$} \\
        \colhead{} & \colhead{} & \colhead{($\rm erg\ cm^{-2}\ s^{-1}$)}  & \colhead{ } }
    \decimalcolnumbers
    \startdata
    2021-2022 & ${2.13\pm0.02}$ & ${(5.75\pm0.19)\times10^{-11}}$ & ${634.6}$ \\
    2022-2023 & ${2.20\pm0.002}$ & ${(7.19\pm0.004)\times10^{-11}}$ & ${868.0}$ \\
    \enddata
    \tablecomments{The column information are as follows: (1) the year of the observation of neutrino events; Col.(2) the best-fit photon index for a power-law model; Col.(3) the integrated $\gamma$-ray flux in the energy range of 0.1-300 GeV; Col.(4) the maximum likelihood test statistic. }
    \end{deluxetable*}

   \begin{deluxetable*}{cccccc}
    \tablenum{2}
    \tablecaption{Results of the Analysis of Swift-XRT Observations of TXS~0506+056\label{tab:messier}}
    \tablewidth{0pt}
    \tablehead{
        \colhead{year} & \colhead{$\rm Exp.$} & \colhead{$N_{\rm H}$} &\colhead{$\rm \Gamma_{x}$} & \colhead{$F_{x}$} & \colhead{$\rm Stat.$} \\
        \colhead{} & \colhead{ks} & \colhead{($10^{21} \ \rm cm^{-2}$)} & \colhead{} & \colhead{($10^{-12} \ \rm erg\ cm^{-2}\ s^{-1}$)} & \colhead{$\rm \chi^{2}$} }
    \decimalcolnumbers
    \startdata
    2021-2022 & ${21.1}$ & ${ 0.96_{-0.27}^{+0.27}}$ & ${1.72_{-0.11}^{+0.11}}$ & ${1.74_{-0.12}^{+0.10}}$  & 35.2 \\
    2022-2023 & ${2.5}$ & ${ 1.16_{-1.03}^{+1.03}}$ & ${2.22_{-0.50}^{+0.50}}$ & ${1.71_{-0.41}^{+0.38}}$ & 0.16 \\
    \enddata
    \tablecomments{The column information are as follows: (1) the year of the observation of neutrino events; Col.(2) the exposure time of X-ray; Col.(3) the neutral hydrogen column density of the extragalactic foreground absorption; Col.(4) the best-fit photon index for a power-law model; Col.(5) the integrated X-ray flux in the energy range of 0.3–10 KeV; Col.(6) the reduced $\chi^{2}$. }
    \end{deluxetable*}
    
\section{MODEL SETUP AND METHOD}
     
     \citet{Ruo_Yu_Liu_2023MNRAS.526.5054L} and \citet{Wang_Ze-Rui_2022} proposed the stochastic dissipation model to explain various observational features of blazars. The model suggests that there are numerous discrete radiation zones along the jet owing to certain dissipation events, such as magnetohydrodynamic instabilities \citep{Giannios_2006A&A, Nalewajko_10.1111}, internal collisions \citep{Deng_wei_2015, Joshi_2011ApJ, Bottcher_2010ApJ...711..445B}, recollimation shocks \citep{Nalewajko2012, Fromm2016}, magnetic reconnections in the jet \citep{Begelman_1998ApJ, McKinney_2009MNRA, Giannios_2009MNRAG, Petropoulou_2016MNRA} and so on, which we do not specify in this work. The probability of a dissipation event occurring at a distance $r$ from the supermassive black hole (SMBH) in unit time and unit length is phenomenologically characterised with a function $p(r)\propto r^{-\alpha}$. The value of $\alpha$ and the injection electron spectral index into each blob can largely determine the spectral shape expected from a blazar.

     The stochastic dissipation model posits that the emission of the blazar jet can be decomposed into two components \citep{Wang_Ze-Rui_2022, Tan2023}. One component is the superposition of emission from numerous but comparatively weak dissipation zones along the entire jet, which is considered to be a quasi-stable, background emission. Another component arises from a strongly dissipating zone in the jet that dominates the emission in the flare state. The historical SED data taken from the SSDC SED builder\footnote{\url{https://tools.ssdc.asi.it/SED/}} representing a long-time average emission of the blazar is then considered as the low-state background emission. Since we here focus on the emission from a flaring zone in the jet to account for the the production of neutrinos, the modelling of the background radiation spectrum is simply characterized with a polynomial function \citep{Wang_Ze-Rui_2022}. \citet{Ruo_Yu_Liu_2023MNRAS.526.5054L} developed a time-dependent model for the low-state background emission, and the result is compatible with the polynomial function approximation.

    To research the neutrino events of TXS 0506+056 in different epochs, we model the broadband SED in each epoch, based on which we can calculate corresponding neutrino detection rate in the epoch. In our model we consider the electron synchrotron radiation, the inverse Compton scattering off the synchrotron radiation field (i.e., synchrotron self-Compton, SSC, \citealt{Harris_2006}) and off the external radiation (i.e., external Compton, EC) of electrons. We also take into account the proton synchrotron radiation, the photopion process, and the Bethe-Heitler process, as well as the emission of pairs generated in the electromagnetic cascade initiated by these processes. 
    
    We assumed that the flaring zone is a blob which has a spherical geometry with a radius $R$, filled with a uniformly entangled magnetic field $B$. The jet’s bulk Lorentz factor $\Gamma$ is assumed to remain constant up to 0.1 pc and we approximate the Doppler factor by $\delta = \Gamma$, $\delta_{0}$ is the Doppler factor at 0.1\ pc. 
    %then we have $\delta = 1/[\Gamma - \left(\Gamma^{2} - 1\right)^{1/2}\cos\theta]$ for a relativistic jet in PKS 1502+036 with a viewing angle of $\theta$, 
    %The Doppler factor for $r>0.1$\,pc can be given by \citep{Potter_2013MN, Wang_Ze-Rui_2022}
    %\begin{equation}
    %\begin{aligned}
    %\delta_{D}(r) = \rm \delta_{D,0}-\frac{\delta_{D,0}-2}{\rm log\left(\frac{100 \ pc}{0.1 \ pc}\right)} \rm log\left(\frac{r}{0.1 \ pc}\right),
    %\end{aligned}
    %\end{equation}
    Considering a truncated conical jet structure and a dissipation zone of radius $R$ comparable to the transverse radius of the jet at certain distance $r$ from the SMBH \citep{Wang_Ze-Rui_2022}, we have
    \begin{equation}
    \begin{aligned}
    R(r) = R_{0}\left(\frac{r}{0.1 \ \rm pc}\right)
    \end{aligned}
    \end{equation}
    where $R_{0}$ is the transverse radius of the jet at 0.1\,pc. Then we can assume that the magnetic luminosity is constant along the jet, which is consistent with results of the VLBA survey \citep{O'Sullivan_2009MN,Sokolovsky_2010arXiv}, the magnetic field strength can be approximated \citep{Wang_Ze-Rui_2022} as a function of $r$
    \begin{equation}
    \begin{aligned}
    B(r) = B_{0}\frac{R_{0}}{R(r)},
    \end{aligned}
    \end{equation}
    where $B_{0}$ is the magnetic field strength of the dissipation zone for r = 0.1 pc.
    
    We assumed that relativistic electrons or protons are injected in the blob with a broken power-law distribution \citep{2010MNRAS.402..497G,2020_Zhenjie} or a power-law distribution, i.e.,
    \begin{equation}
    \begin{aligned}
    Q_{\rm e(p)}(\gamma_{\rm e(p)})=\ Q_{\rm e(p),0}\gamma_{\rm e(p)}^{-n_{\rm e(p),1}}\left[1+\left(\frac{\gamma_{\rm e(p)}}{\gamma_{\rm e(p),b}}\right)^{\left(n_{\rm e(p),2}-n_{\rm e(p),1}\right)}\right]^{-1},
    \end{aligned}
    \end{equation}
    \begin{equation}
    \begin{aligned}
    Q_{\rm e(p)}(\gamma_{\rm e(p)})&=\ Q_{\rm e(p),0}\gamma_{\rm e(p)}^{-n_{\rm e(p)}},
    \end{aligned}
    \end{equation}
    Here $\gamma_{\rm e(p)min}$ and $\gamma_{\rm e(p)max}$ are the minimum and maximum Lorentz factor of electrons or protons respectively, $\gamma_{\rm e(p),b}$ is the break Lorentz factor of electrons or protons, $Q_{\rm e(p),0}$ is the normalization, $n_{\rm e(p),1}$ and $n_{\rm e(p),2}$ represent the low-energy slope and the high-energy slope of the broken power-law distribution respectively. $Q_{\rm e(p),0}$ can be obtained from $\int Q_{\rm e(p)}\gamma_{\rm e(p)}m_{\rm e(p)}c^{2}d\gamma_{\rm e(p)} = L_{\rm e(p),inj}/\left(4/3\pi R^{3}\right)$ where $c$ is the speed of light and $m_{\rm e(p)}$
    is the rest mass of electron or proton, $L_{\rm e(p),inj}$ is the injection luminosity of electrons or protons. The steady-state density distribution of electrons(protons) can be written as
    \begin{equation}
    N_{\rm e(p)}(\gamma_{\rm e(p)}) = Q_{\rm e(p)}(\gamma_{\rm e(p)})t_{\rm e(p)},
    \end{equation}
    where $t_{\rm e(p)}$ is the minimum value among $t_{\rm cool}$ and $t_{\rm e(p),dyn}$. $t_{\rm cool} = 3m_{\rm e}c / \left({4\left(U_{\rm B}+\kappa_{\rm KN}U_{\rm ph}\right)\sigma_{\rm T}\gamma_{\rm e}}\right)$ is the radiative cooling timescale of electron. $U_{\rm B} = B^{2}/8\pi$ is the energy density of the magnetic field and $U_{\rm ph}$ is the energy density of the soft photons. Here $\sigma_{T}$ is the Thomson scattering cross section and $\kappa_{\rm KN}$ is a factor accounting for the Klein-Nishina effect\citep{2005MNRAS.363..954M}. For protons, we need to include the energy loss caused by the photopion production and the Bethe-Heitler (BH) pair production when calculating the cooling timescale. These two processes will be further described later. $t_{\rm e(p),dyn} = R/c$ is the electron(proton) dynamical timescale, here we assumed the radius of the blob is $R = r\sin \theta$, $r$ is the distance between the blob and the black hole. In this work all timescales are evaluated in the comoving frame of the blob. 
    
    Under the unified model, the synchrotron, SSC and EC radiations are calculated following \citet{Wang_Ze-Rui_2022}. If the blob (flaring zone) is close to the supermassive black hole (SMBH), the radiation of the broad line region (BLR) and the dust torus (DT) could enter the blob with a significant amplification of the energy density due to the Doppler effect. The radiation field of BLR is assumed to be a blackbody emission peaking at a frequency $2 \times 10^{15}\Gamma \rm \ Hz$ \citep{2008MNRAS.386..945T}, and the radiation field of DT is assumed to be a blackbody emission peaking at a frequency $3\times10^{13}\Gamma \rm \ Hz$ \citep{2007ApJ...660..117C} in the jet comoving frame, respectively. 
    %In the EC process, if the energy density of the BLR dominates the EC process is named EC/BLR, otherwise it is named EC/DT. 
    The energy density of external photons from the BLR and DT can be approximated \citep{2012ApJ...754..114H} by
    \begin{equation}
    u_{\rm BLR} = \frac{\eta_{\rm BLR}\Gamma^{2}L_{\rm d}}{3\pi r_{\rm BLR}^{2}c[1+\left(r/r_{\rm BLR}\right)^{3}]}, 
    \end{equation}
    and
    \begin{equation}
    u_{\rm DT} = \frac{\eta_{\rm DT}\Gamma^{2}L_{\rm d}}{3\pi r_{\rm DT}^{2}c[1+\left(r/r_{\rm DT}\right)^{4}]}, 
    \end{equation}
    where $\eta_{\rm BLR} = \eta_{\rm DT} = 0.1$ are the fractions of the disk luminosity $L_{\rm d}$ reprocessed into the BLR and DT radiation, respectively.
    We assumed the characteristic distance of BLR in the AGN frame is $r_{\rm BLR} = 0.1\left(L_{\rm d}/10^{46}\rm erg \ s^{-1}\right)^{1/2} \rm pc$ and the characteristic distance of DT is $r_{\rm DT} = 2.5\left(L_{\rm d}/10^{46}\rm erg \ s^{-1}\right)^{1/2}\,$pc. The disk luminosity $L_{\rm d}$ is assumed to be $5\times10^{44}\rm erg\ s^{-1}$ \citep{Xue_2021}.
    Meanwhile, the X-ray photons emitted by the corona surrounding the accretion disk \citep{Heckman_2014ARA&A} are also considered in our model, if the blob is near or at the jet base with a distance comparable to only a few times larger than the Schwarzschild radius ($r_{\rm Sch}\sim 10^{14}(M_{\rm SMBH}/3\times10^{8}M_{\odot})\rm cm$) of the central SMBH, the X-ray photons would interact with gamma-ray photons emitted by the blob \citep{Righi_2019MNR}. We assumed the emission of the corona has a power-law spectrum \citep{Ricci_2018MNR, Xue_2021}
    \begin{equation}
    L(E) = L_{1 \ \rm keV}(E/1 \ \rm keV)^{1-\alpha}, \ 0.1 \ \rm keV < E < 100 \ \rm keV,
    \end{equation}
    here we simply assume that its spectral index $\alpha = 1$ and $L_{1 \ \rm keV}$ is set as $5\times10^{43}\rm erg\ s^{-1}$ for TXS~0506+056 \citep{Xue_2021}. The corresponding energy density in the comoving frame can be written as
    \begin{equation}
    u_{\rm corona} (r) = \frac{\Gamma^{2}\int L(E)/EdE}{4\pi r^{2}c},
    \end{equation}
    where $r$ is the distance between the central black hole and the dissipation region (or the blob). Eq.~(10) is valid for $r\gg r_{\rm Sch}$, where as for $r<10r_{\rm Sch}$ we assume $u_{\rm corona}$=$u_{\rm corona}(r=10r_{\rm Sch})$. 

    Note that the relativistic protons can also interact with these radiation field via the photopion process and Bethe–Heitler process. The photopion process is the production of pions in proton-photon interactions:
    \begin{equation}
    \begin{aligned}
    p& + \gamma \rightarrow p^{\prime} + \pi^{0}, \\
    p& + \gamma \rightarrow n + \pi^{+}, \\
    p& + \gamma \rightarrow p^{\prime} + \pi^{+} + \pi^{-}, \
    \end{aligned}
    \end{equation}
    via which unstable pions are produced and pions will decay into
    \begin{equation}
    \begin{aligned}
    \pi^{0}& \rightarrow 2\gamma, \\
    \pi^{+} \rightarrow \mu^{+} + \nu_{\mu}& \rightarrow e^{+} + \nu_{e} + \Bar{\nu_{\mu}} + \nu_{\mu},\\
    \pi^{-} \rightarrow \mu^{-} + \Bar{\nu_{\mu}}& \rightarrow e^{-} + \bar\nu_{e} + \nu_{\mu} + \Bar{\nu_{\mu}},\
    \end{aligned}
    \end{equation}
    Another important process is the Bethe–Heitler process, this process will lead to the production of electron/positron pairs, i.e.,
    \begin{equation}
    \begin{aligned}
    p + \gamma \rightarrow p^{\prime} + e^{+} + e^{-}, \
    \end{aligned}
    \end{equation}
    \citet{2008PhRvD..78c4013K} developed a semi-analytical method to calculate spectra of secondary particles in these two processes and we followed the method in \cite{2008PhRvD..78c4013K}. The cooling timescales of both electrons and protons via the aforementioned processes in the model are shown in Figure 1 for references.
    
    The maximum Lorentz factor of protons in the emission region can be approximated by equating the acceleration and the cooling or dynamical timescales
    \begin{equation}
    \begin{aligned}
    t_{\rm acc} = {\rm min}\left\{t_{p\gamma},t_{\rm BH},t_{\rm dyn},t_{\rm p,syn}\right\},
    \end{aligned}
    \label{f6} 
    \end{equation}
    The acceleration timescale can be evaluated by \citep{Rieger_2007Ap&SS}
    \begin{equation}
    \begin{aligned}
    t_{\rm acc}\simeq\frac{\alpha r_{\rm L}}{c}\simeq\frac{\alpha\gamma_{\rm p}m_{\rm p}c}{eB},
    \end{aligned}
    \end{equation}
    where $\alpha(\geq 1)$ is the parameter characterising the particle acceleration efficiency ($\alpha=1$ being the most efficient case), which depends on the spectrum of magnetic turbulence and on the velocity of the scattering center in the case of Fermi-type acceleration \citep{2011IAUS..275...59S}. Here we employ $\alpha = 10$ in our model. The kinetic luminosity of the nonthermal particles and magnetic field can be written as \citep{Celotti_2008}
    \begin{equation}
    L_{k,i} = \pi R^{2}\Gamma^{2}cU_{i}, 
    \end{equation}
    where $U_{i}$ is the energy density of nonthermal particles or magnetic field, here i = $\left\{\rm e ,\rm p, \rm B \right\}$, represents electrons, protons and magnetic field respectively. 
    
    \begin{figure}[ht!]
    \plotone{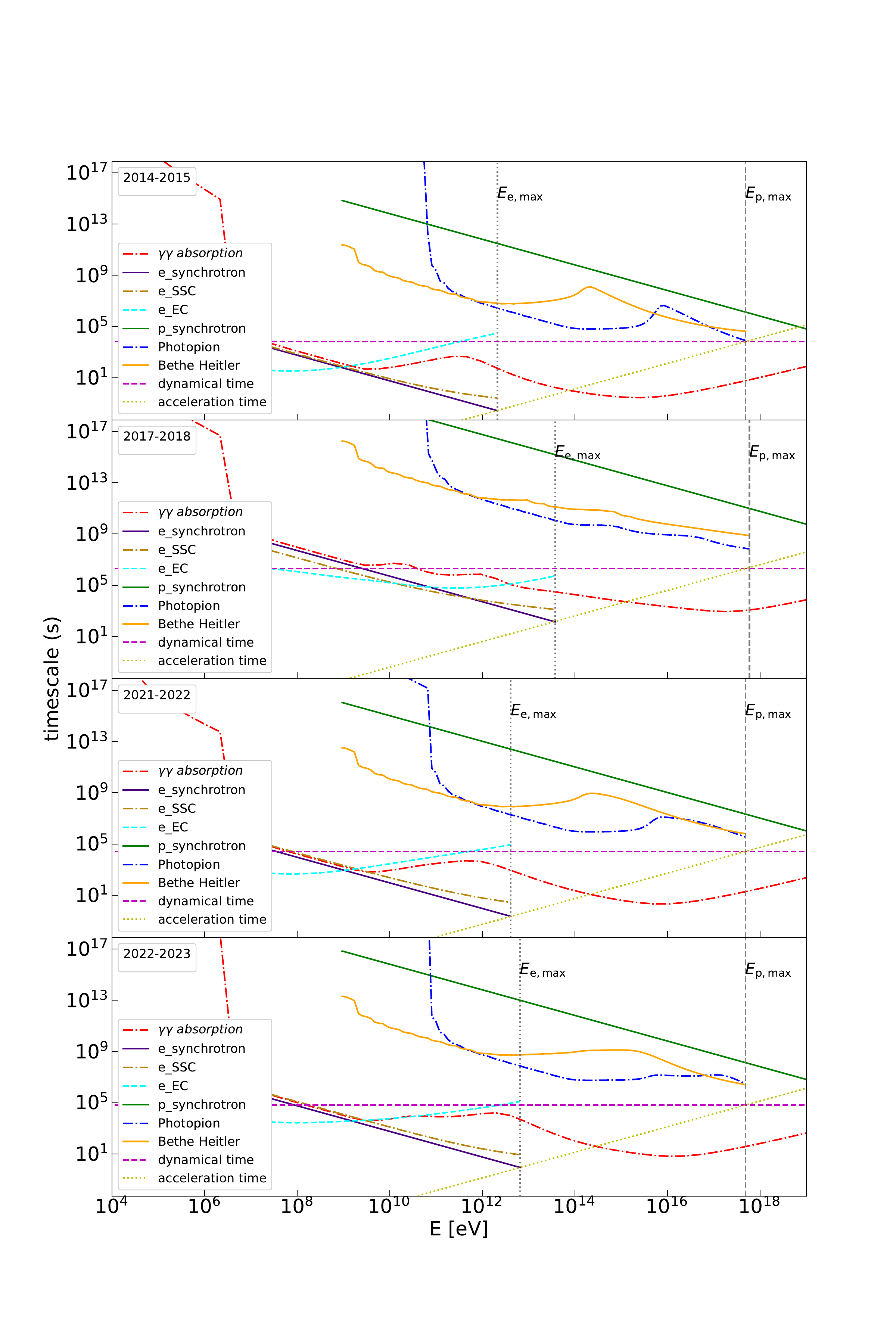}
    \caption{The cooling timescale of protons and electrons in different epochs, as well as the opacity of the intrinsic $\gamma\gamma$ absorption of the radiation zone. The horizontal axis $\rm eV$ represents ernergy of protons for synchrotron, photopion and Bethe-Heitler process in the comoving frames, and also represents ernergy of electrons for synchrotron, SSC and EC process in the comoving frames. In addition, we also plot the optical depth of photons(the red dot-dashed line) in the figure. The vertical grey dashed lines show the maximum electron and proton energy allowed by the $t_{\rm acc} = $\rm min$ \left\{t_{\rm cool},t_{\rm e(p),dyn}\right\}$. Parameters of the model are described in Table 3, respectively.
    \label{fig:general}}
    \end{figure}
    
\section{SED Modeling and results}

    In this section, we use the stochastic dissipation model to reproduce the mutiwavelength emission of TXS~0506+056 in the four epochs with neutrino detection. For the photopion process and the BH process,
    radiation fields of the blazar serve as targets, including the emission of the BLR, the dusty torus and the corona, the radiation of primary electrons, as well as the radiation of secondary pairs developed in the electromagnetic cascade initiated by the two processes. The BH process and photopion process including the cascade initiated by these processes are calculated following the method shown in \citep{2008PhRvD..78c4013K, Bottcher_2013}.
    % Fig 2 - 14-15.pdf    
    \begin{figure}[ht!]
    \plotone{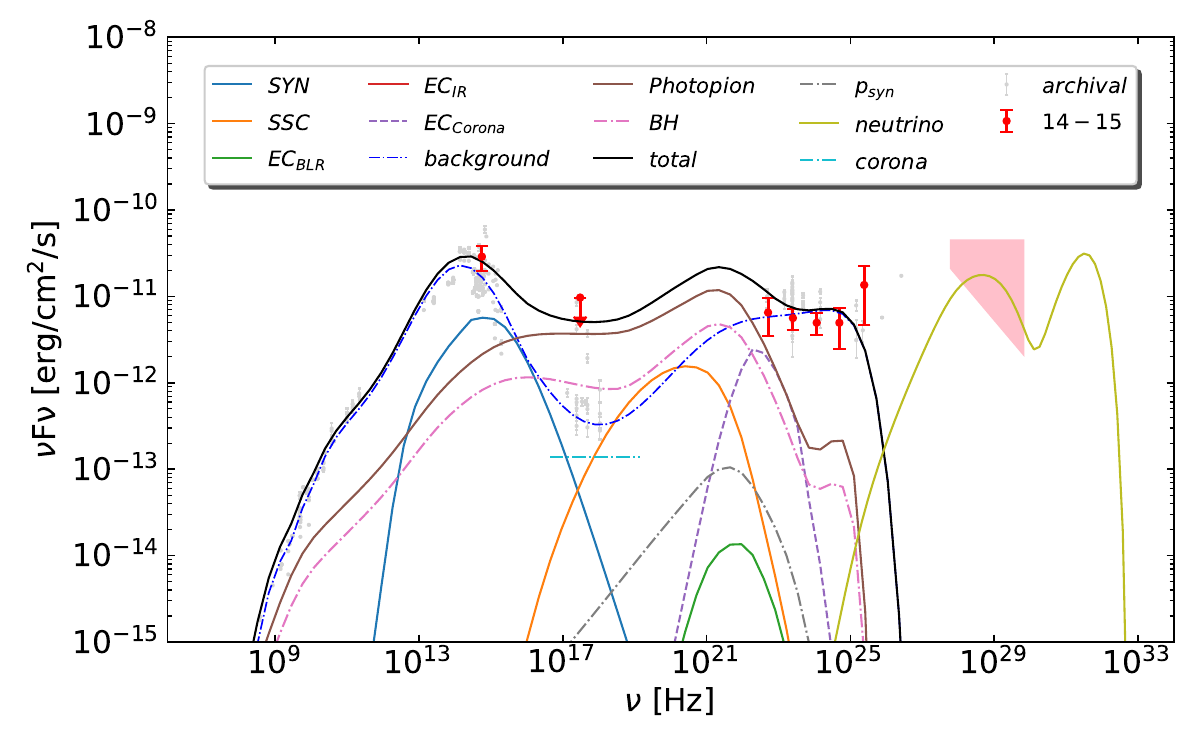}
    \caption{Fitting to the broadband SED of TXS 0506+056 in 2014-2015. The red data are from \citet{Rodrigues_2019ApJ}, the gray data points are historical archival data. The pink bow tie shows the muon-neutrino ﬂux as measured during the 14–15 ﬂare \citep{2018Sci...361..147I}.}
    \label{fig:general}
    \end{figure}
    
    To reduce the number of free parameters, value of $\gamma_{\rm e,min}$ in different neutrino events of TXS~0506+056 are fixed to 50. Our model has four parameters ($R_{0}$, $B_{0}$, $n_{\rm e,1}$, and $\delta_{0}$) that are common among different zone, and four parameters ( $L_{\rm e}^{\rm inj}$, $n_{\rm e,2}$, $\gamma_{\rm break}$ and $r$) that are unique to each dissipation zone. $\gamma_{\rm e,max}$ and $\gamma_{\rm p,max}$ are obtained by equating the acceleration and the cooling or dynamical timescales. Then we assumed protons are injected in the blob with a power-law distribution. The spectral index $n_{\rm p}$ and the minimum Lorentz factor $\gamma_{\rm p,min}$ of the proton distribution are fixed to 2 and 1 respectively, while the maximum injection luminosity of protons is obtained when the pair cascade emission saturates the X-ray or $\gamma$-ray data. We have a total number of 20 parameters in the model, with 4 common for all four epochs and 4 parameters that are separate for each of them.  The best-fit and uncertainty of the model parameters are obtained via the Markov Chain Monte Carlo (MCMC) method with the PYTHON package EMCEE \citep{emcee_2013}.
    
\subsection{Epoch I: 2014-2015}
    
    The SED fitting result is shown in Figure 2 and the derived parameters are listed in Table 3. The best-fit blob distance is $3\times 10^{-4}$\,pc, or approximately $10\,r_{\rm rsc}$, from the SMBH, which is located within the corona.  
    
    At this distance the X-ray photons emitted by the hypothetical corona surrounding the accretion disk is very dense and absorb the $\gamma$-ray photons \citep{Kamraj_2018ApJ...866..124K, Righi_2019MNR, Ricci_2018MNR} efficiently, so the observed $\gamma$-ray emission is mainly contributed by the background emission. This explains why the blazar is in the gamma-ray quiescent state during the neutrino outburst. Moreover, the X-ray corona of SMBH could be a promising neutrino production site provided that protons are accelerated in the region. The number of predicted muon neutrino events detected by IceCube during half a year is 8.2 based on the best-fit parameters, which is basically consistent with IceCube's measurement \citep{2018Sci...361..147I}. 
    
    \citet{Xue_2021} used a two-zone radiation model of blazars to explain the broadband SED of TXS~0506+056 during 2014-2015, the derived distance between the black hole and the inner blob is also approximately several times the Schwarzschild radius. In their model, the magnetic field of the inner blob is taken to be a free parameter and is only 2\,G. The value is much smaller than the best-fit value 83.3\,G in our work, which is based on a $B\propto r^{-1}$ relation and a simultaneous fitting to SEDs of other three epochs. The strong magnetic field makes synchrotron radiation comparably efficient as the EC process, or even the dominant radiation process for electrons above GeV energies considering the KN effect (see Figure \ref{fig:general}). The synchrotron radiation of pairs developed in the EM cascade creates a bump around MeV band in the SED. This is consistent with the results in some previous literature for the neutrino outburst \citep{Xue_2021} and could be hopefully detected by the next-generation MeV instrument as a test for the model. At GeV band, the high opacity of the $\gamma\gamma$ absorption due to the X-ray corona photon is very high for the blob. Therefore, the blazar's GeV emission is dominated by the background component, which is consistent with observation.
    
    The kinetic luminosity of the magnetic field, electrons and protons in the blob are $1.09\times10^{44}~\rm erg/s$, $3.09\times10^{42}~\rm erg/s$ and $2.6\times10^{46}~\rm erg/s$ respectively. The SMBH mass of TXS~0506+056 is $3\times10^{8}M_{\odot}$ \citep{Padovani_2019MNRA}, the corresponding Eddington luminosity of TXS~0506+056 is $L_{\rm edd}\approx3.01\times10^{46}(M_{\rm BH}/10^{8.5}M_{\odot})~\rm erg/s$, so the total kinetic luminosity is smaller than the Eddington luminosity.

\subsection{Epoch II: 2017-2018}

    The best-fit distance of the blob for this epoch is located at 0.09\,pc from the SMBH and is the largest one among the four epochs. This distance is far beyond the size of the corona and comparable to the typical radius of the BLR. The external radiation field in this epoch is the BLR radiation and the DT radiation. The absorption of gamma-ray photons in the blob is not as important as that for Epoch I. The corresponding magnetic field in this blob is 0.28\,G, which is similar to those found in many previous studies for the multi-messenger emission of IC-170922A \citep{Xue_2019, Liu_2019, Das_2022, Cerruti_2019, 2019NatAs...3...88G, Gasparyan_2021}. 

    The EM radiation of the blob is important in this epoch, where the optical flux is comparable to that of the background emission and the gamma-ray emission dominates. This is consistent with the observed flux enhancement in the optical band and the gamma-ray flare around the arrival time of IC-170922A \citep{2018Sci...361.1378I}. The neutrino emission is produced by the photopion process between protons and BLR radiation. The number of predicted muon neutrino events detected by IceCube during half a year is approximate 0.07 based on the best-fit parameters, which is an acceptable value.
    
    The SED fitting result is shown in Figure 3 and the derived parameters are listed in Table 3. The maximum Lorentz factor of electrons and protons are $7.2\times10^{7}$ and $6.2\times10^{8}$ respectively. The kinetic luminosity of the magnetic field, electrons and protons in the blob are $1.09\times10^{44}~\rm erg/s$, $3.56\times10^{44}~\rm erg/s$ and $2.61\times10^{46}~\rm erg/s$ respectively. The total luminosity is close but still below the Eddington luminosity of the SMBH.
    
    % Fig 3 - 2017-2018.pdf
    \begin{figure}[ht!]
    \plotone{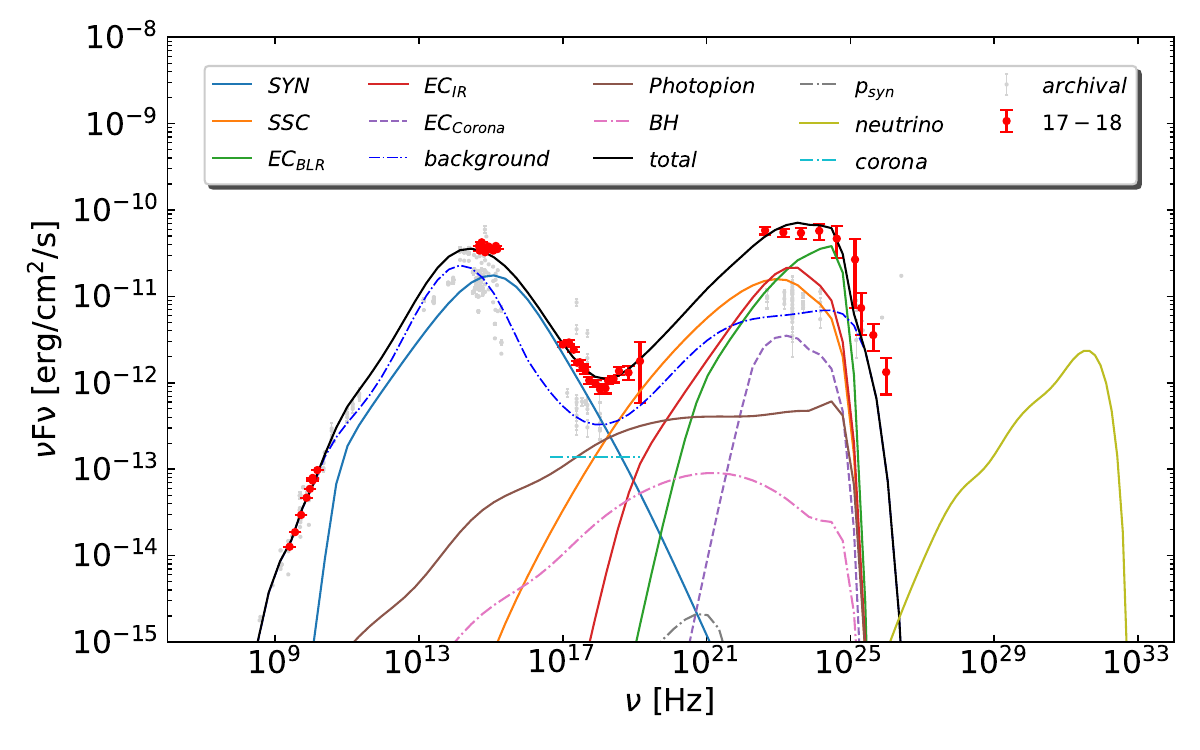}
    \caption{Fitting to the broadband SED of TXS 0506+056 in 2017-2018. The red data are from \citet{2018Sci...361.1378I}, the gray data points are from archival.\label{fig:general}}
    \end{figure}

    % Fig 4 - 2021-2022.pdf
    \begin{figure}[ht!]
    \plotone{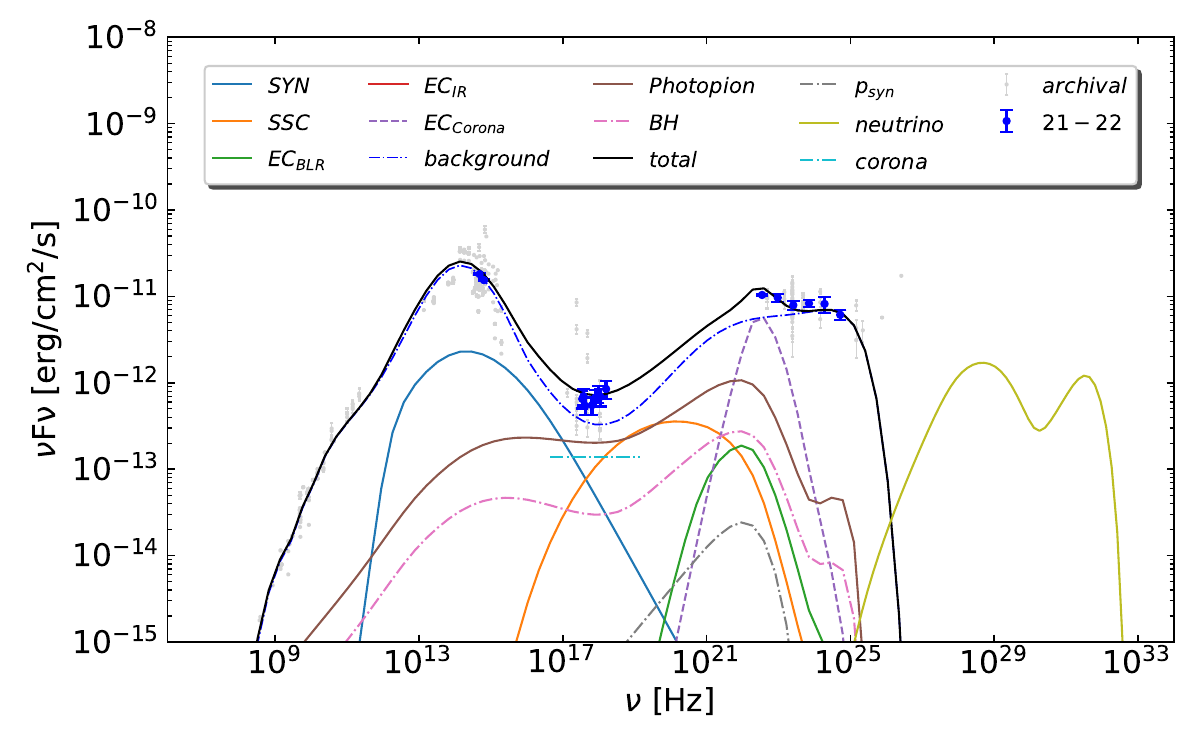}
    \caption{Fitting to the broadband SED of TXS 0506+056 in 2021-2022. The blue data are from this work, the gray data points are from archival.\label{fig:general}}
    \end{figure}

    % Fig 5 - 2022-2023.pdf
    \begin{figure}[ht!]
    \plotone{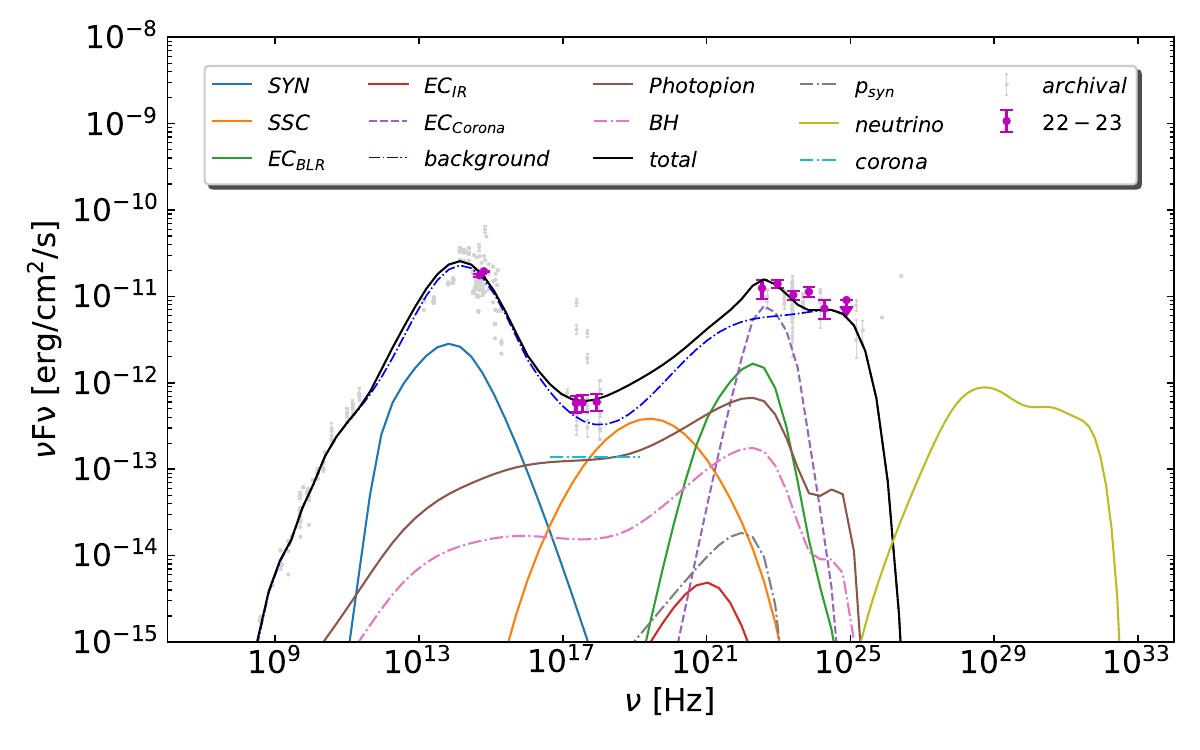}
    \caption{Fitting to the broadband SED of TXS 0506+056 in 2022-2023. The purple data are from this work, the gray data points are from archival.\label{fig:general}}
    \end{figure}

\subsection{Epoch III: 2021-2022}

    The SED fitting result is shown in Figure 4 and the derived parameters are listed in Table 3.  The magnetic field of the blob is approximately 20.8 G in this case, the distance between the black hole and the blob (flaring zone) is approximately 0.0012 pc. This value is 4 times the distance of the flaring blob in Epoch I. The predicted number of neutrino events, if assuming the muon neutrino effective area of IceCube, is 0.73 for half a year. Note that this neutrino event, GVD-210418CA, is a cascade event detected by Baikal-GVD. The effective area of Baikal-GVD for cascade event is several times smaller than that of IceCube for the track channel \citep{Baikal-GVD_2021}, so we expect the detection rate of neutrino in this epoch by Baikal-GVD is at the order of $\sim 0.1$, which is not unreasonable for the detection. From the perspective of SED fitting, as shown in Figure 4 and Figure 2, we can see that the observed X-ray flux in Epoch III is lower than the X-ray flux upper limit in Epoch I. This leads to a restraint on neutrino flux in Epoch III than that in Epoch I, because otherwise the observed X-ray flux would be overshot by the cascade emission initiated by proton–photon interactions. As shown in Table 3, the total kinetic luminosity is $8.20\times10^{45}~\rm erg/s$. This value is about three times smaller than that in Epoch I. Also, given the larger distance, the energy density of the X-ray corona has decreased by one order of magnitude. Therefore the photopion production efficiency decreases, as can be seen in Figure 1 by comparing the ratio of the cooling timescale of photopion to the dynamical timescale in this two Epochs.
    
    On the other hand, where the X-ray photons emitted by the corona is still very intense at this distance and absorb the $\gamma$-ray photon above GeV emitted by the blob efficiently \citep{Kamraj_2018ApJ...866..124K, Righi_2019MNR}. As a result, the blob only has a comparable contribution to the gamma-ray flux around 0.1\,GeV, leading to a slightly soft gamma-ray spectrum. The blazar thus appears in a low state in the GeV band during this epoch, which is consistent with the observation.
    
\subsection{Epoch IV: 2022-2023}

    The SED in Epoch IV is similar to that in Epoch III. The X-ray flux in this epoch is slightly lower than that in Epoch III while the gamma-ray flux around 0.1\,GeV is a little bit higher. As a result, the best-fit distance of the blob is a few times farther away from the SMBH (0.003\,pc from the SMBH), resulting in a slightly lower photopion efficiency and gamma-ray opacity. Comparing with the Epoch III, the blob's contribution at 0.1\,GeV is a little bit higher but the emission in the Fermi-LAT band is still dominated by the background component. The lower X-ray flux leads to a lower neutrino flux, yielding 0.41 muon neutrino events in half a year. The kinetic luminosity of the magnetic field, electrons and protons in the blob are $1.09\times10^{44}~\rm erg/s$, $1.82\times10^{43}~\rm erg/s$ and $1.12\times10^{46}~\rm erg/s$ respectively. The SED fitting result is shown in Figure 5 and the derived parameters are listed in Table 3.

    \begin{deluxetable*}{ccccccccccccc}
    \tablenum{3}
    \tablecaption{The physical parameters derived from the unified models fits \label{tab:messier}}
    \tablewidth{0pt}
    \tablehead{
        \colhead{year} & \colhead{$\log R_{0}$} & \colhead{$\delta_{0}$} &\colhead{$B_{0}$} & \colhead{$p_{1}$} & \colhead{$\log r$} & \colhead{$p_{2}$} & \colhead{$\log \gamma_{break}$} & \colhead{$\log L_{\rm e,inj}$} &\colhead{$B$} &\colhead{$L_{k,\rm B}$} & \colhead{$L_{k,\rm p+e}$} & \colhead{$N_{\nu_\mu}$} \\
        \colhead{} & \colhead{(cm)} & \colhead{} & \colhead{(G)} & \colhead{} & \colhead{(pc)} & \colhead{} & \colhead{} &\colhead{($\rm erg \ \rm s^{-1}$)} &\colhead{(G)} &\colhead{($\rm erg \ \rm s^{-1}$)} & \colhead{($\rm erg \ \rm s^{-1}$)}}
    \decimalcolnumbers
    \startdata
    2014-2015 & \multirow{4}{*}{16.81$_{-0.12}^{+0.16}$} & \multirow{4}{*}{10.73$_{-1.3}^{+1.0}$} & \multirow{4}{*}{0.25$_{-0.07}^{+0.07}$} &  \multirow{4}{*}{1.39$_{-0.23}^{+0.22}$} & -3.52$_{-0.27}^{+0.32}$ & 4.62$_{-0.31}^{+0.24}$ & 2.73$_{-0.30}^{+0.26}$ & 42.87$_{-0.41}^{+0.42}$ &83.3 &1.09e44 & 2.60e46 & 8.2\\
        2017-2018 & ${}$ & ${}$ & ${}$ & ${}$ & -1.05$_{-0.08}^{+0.08}$ & 3.86$_{-0.21}^{+0.29}$ & 4.10$_{-0.24}^{+0.17}$ & 43.49$_{-0.40}^{+0.32}$ &0.28 & 1.09e44 & 2.65e46 & 0.07 \\
            2021-2022 & ${}$ & ${}$ & ${}$ & ${}$ & -2.93$_{-0.30}^{+0.31}$ & 3.57$_{-0.39}^{+0.26}$ & 2.51$_{-0.21}^{+0.33}$ & 42.62$_{-0.25}^{+0.30}$ &20.8 & 1.09e44 & 8.20e45 & 0.73$^a$ \\
                2022-2023 & ${}$ & ${}$ & ${}$ & ${}$ & -2.53$_{-0.32}^{+0.33}$ & 4.63$_{-0.22}^{+0.20}$ & 2.81$_{-0.35}^{+0.30}$ & 42.82$_{-0.41}^{+0.40}$ &8.3 & 1.09e44 & 1.12e46 & 0.41 \\
    \enddata
    \tablecomments{The column information are as follows: (1) the year of the observation of neutrino events; Col.(2) the transverse radius of the jet at 0.1 pc; Col.(3) the Doppler factor at 0.1 pc; Col.(4) the magnetic field strength of the dissipation zone for r = 0.1 pc; Col.(5) low energy spectral index of electron; Col.(6) the distance between black hole and blob (flaring zone); Col.(7) high energy spectral index of electrons; Col.(8) the break Lorentz factor of electrons; Col.(9) the injection luminosity of electrons; Col.(10) the magnetic field strength of the blob; Col.(11) the luminosity of magnetic field of the blob; Col.(12) the total luminosity of electrons and protons of the blob; Col.(13) the numbers of neutrino events (during half a year). We assumed protons are injected in the blob with a power-law distribution. The spectral index and the minimum Lorentz factor of the proton distribution are fixed to 2 and 1 respectively, while the maximum Lorentz factor of the proton is obtained by equating the acceleration and the cooling or dynamical timescales. The maximum injection luminosity of protons is obtained when the pair cascade emission saturates the X-ray or $\gamma$-ray data.}.\\
    $a$: calculated with IceCube's effective area for the track channel, see Section 4.3 for more discussion.
    \end{deluxetable*}

\section{Discussion and Summary}

    In this work, we applied the stochastic dissipation model to simultaneously explain neutrino events and broadband SED of TXS~0506+056 detected in four different epochs, corresponding to the 2014-2015 neutrino outburst and the arrival time of three single muon neutrino event IC-170922A, GVD-210418CA, and IC-220918A. In this model, the emission of the jet is composed of two emission component \citep{Wang_Ze-Rui_2022}. One emission component is the superposition of emission from numerous but comparatively weak dissipation zones along the entire jet, which is considered to be the same in all four epochs. Another component arises from a compact flaring zone (i.e., a blob with intense particle injection) that could form at any position in the jet randomly. In addition to radiation of electrons considered by \citet{Wang_Ze-Rui_2022}, we also assume protons are accelerated in the flaring zone and undergo synchrotron radiation and proton–photon interactions (including the Bethe–Heitler process and the photopion process), as well as the emission of pairs generated in the EM cascade initiated by these processes. Some of properties of the flaring zones in different epochs such as the magnetic field and the Doppler factor are related to each other, while the injection luminosity of relativistic electrons and protons and the distance from the SMBH are independent. In particular, the distance of the flaring zone determines the EM environment it is located and leads to different radiation properties. 
    
    The best-fit distance of the flaring zones in all four epochs are small, where the one during the Epoch I (2014-2015), Epoch III (2021-2022), Epoch IV (2022-2023) are close to the X-ray corona and the one during Epoch II (2017-2018) is beyond the X-ray corona but still within the BLR. These external radiation fields provide promising target photons for neutrino production.  
    Compared to conventional one-zone models, this model can yield a higher neutrino flux with a more reasonable detection probability of these neutrino events, i.e., 8.2, 0.07, 0.73 ($\sim 0.1$ for Baikal-GVD), and 0.41 in half a year for Epoch I-IV respectively, while the required kinetic luminosity of the jet is below the Eddington luminosity. Then, we may expect an overall probability of detecting the three single-neutrino event in Epoch II, III, and IV, to be $0.07\times 0.1\times 0.41\approx 0.003$, which is marginally within $3\sigma$ confidence level and not too small to be unreasonable. 
    
    We note that the neutrino flux is positively correlated to the X-ray flux of this blazar, which is similar to that predicted in conventional one-zone models. This is because the co-produced EM particles in the photopion production can initiate EM cascades and deposit energy in lower energy band via synchrotron radiation of electron/positron pairs generated in the cascade, while the X-ray flux radiated by primary electrons accelerated in the jet is weak for this blazar. In fact, the main observational constraints on the neutrino flux is from the measured X-ray flux. This also explains why the predicted neutrino detection rate in this model is much higher with respect to that in conventional one-zone models: the low-energy synchrotron bump in the SED is not fully contributed by the neutrino emission zone, so a weaker magnetic field can be employed in the neutrino emission zone which also reduces the cascade emission at the X-ray band. This is similar to reason of a higher neutrino flux in the two-zone model as discussed by \citet{Xue_2019}.

%    As mentioned the number of predicted muon neutrino events detected by IceCube during 2021-2022 is approximately 0.73. It is noted that this value is obtained by the effective area of IceCube, but the observed neutrino during 2021-2022 is observed by Baikal-GVD \citep{Baikal-GVD_2021}, the effective area of Baikal-GVD is several times smaller than the effective area of IceCube \citep{Baikal-GVD_2021}, so the corresponding number of predicted muon neutrino events detected by Baikal-GVD is also several times smaller than 0.73, and is also a acceptable value.  
    
    To conclude, the blazar TXS~0506+056 is a candidate of high-energy neutrino emitter. In addition to the 2014-2015 neutrino outburst and IC-170922A as reported by IceCube several years ago, two additional high-energy muon neutrino events, GVD-210418CA and IC-220918A, are reported recently in spatial association with the blazar \citep{Baikal-GVD_2021, Becker_2022arXiv}. If the neutrino events in all four epochs truly originate from TXS~0506+056, the conventional one-zone lepto-hadronic model, with which IC-170922A was explained as a lucky detection \citep{2018ApJ...864...84K, Cerruti_2019, 2019NatAs...3...88G}, can be ruled out, because the probability of detecting all these neutrino events from the blazar under the one-zone model would be unreasonably low. We attempt to extend the stochastic dissipation model by including hadronic emissions and apply the model to this blazar. We found that SEDs of TXS~0506+056 in all four epochs with neutrino detection can be simultaneously explained and the predicted neutrino event are reasonable. The model supports blazars to be efficient high-energy neutrino emitters and sheds some light on the radiation models of blazars.  

%    For four neutrino events, the magnetic field of the blob are approximately 83.3 G, 0.28 G, 20.8 G and 8.3 G respectively. These values are also roughly comparable to the values at these distances predicted in \citet{O'Sullivan_2009MN}. The required total kinetic luminosity is approximately $2.61\times10^{46}~\rm erg/s$, $2.66\times10^{46}~\rm erg/s$, $8.31\times10^{45}~\rm erg/s$ and $1.13\times10^{46}~\rm erg/s$ respectively in our work, these values are smaller than the Eddington luminosity $L_{\rm Edd} \approx 4\times 10^{46}\left(M_{\rm BH}/10^{8.5} M_{\odot} \right) \rm erg\ s^{-1}$, with $M_{\rm BH}$ $\approx$ $10^{8.5}$ $M_{\odot}$ being the SMBH mass of this object \citep{Padovani_2019MNRA}. In \citet{Tchekhovskoy_2011MNRAS}, the numerical simulations results also show that when the black hole is rapidly spinning, magnetically arrested accretion can efficiently generate jets. In the process of magnetically arrested accretion, the energy of the magnetic field is converted into the kinetic energy of the jets, leading to a highly efficient generation of jets. 

\section*{acknowledgments}

We sincerely thank the anonymous referee for the constructive comments and helpful suggestions. JFW was supported by the National Key R\&D Program of China (Grant No. 2023YFA1607904). ZJW and JFW acknowledge support by the National Science Foundation of China (NSFC) grants No. 12033004 and 12221003. RYL acknowledges NSFC grant No.~U2031105. This research has made use of the NASA/IPAC Extragalactic Database, which is funded by the National Aeronautics and Space Administration and operated by the California Institute of Technology.

\vspace{5mm}
%\facilities{HST(STIS), Swift(XRT and UVOT), AAVSO, CTIO:1.3m,
%$CTIO:1.5m,CXO}



\bibliography{sample631}

\begin{thebibliography}{}
\expandafter\ifx\csname natexlab\endcsname\relax\def\natexlab#1{#1}\fi
\providecommand{\url}[1]{\href{#1}{#1}}
\providecommand{\dodoi}[1]{doi:~\href{http://doi.org/#1}{\nolinkurl{#1}}}
\providecommand{\doeprint}[1]{\href{http://ascl.net/#1}{\nolinkurl{http://ascl.net/#1}}}
\providecommand{\doarXiv}[1]{\href{https://arxiv.org/abs/#1}{\nolinkurl{https://arxiv.org/abs/#1}}}

\bibitem[{Aartsen {et~al.}(2020)Aartsen, Ackermann, Adams, Aguilar, Ahlers, Ahrens, Alispach, Andeen, Anderson, Ansseau, Anton, Argüelles, Auffenberg, \& et~al.}]{IceCube_2020}
Aartsen, M., Ackermann, M., Adams, J., {et~al.} 2020, Physical Review Letters, 124, \dodoi{10.1103/physrevlett.124.051103}

\bibitem[{{Aartsen} {et~al.}(2013){Aartsen}, {Abbasi}, {Abdou}, {Ackermann}, {Adams}, {Aguilar}, {Ahlers}, {Altmann}, {Auffenberg}, {Bai}, {Baker}, {Barwick}, {Baum}, {Bay}, {Beatty}, {Bechet}, {Becker Tjus}, {Becker}, {Bell}, {Benabderrahmane}, {BenZvi}, {Berdermann}, {Berghaus}, {Berley}, {Bernardini}, {Bernhard}, {Bertrand}, {Besson}, {Binder}, {Bindig}, {Bissok}, {Blaufuss}, {Blumenthal}, {Boersma}, {Bohaichuk}, {Bohm}, {Bose}, {B{\"o}ser}, {Botner}, {Brayeur}, {Bretz}, {Brown}, {Bruijn}, {Brunner}, {Carson}, {Casey}, {Casier}, {Chirkin}, {Christov}, {Christy}, {Clark}, {Clevermann}, {Coenders}, {Cohen}, {Cowen}, {Cruz Silva}, {Danninger}, {Daughhetee}, {Davis}, {De Clercq}, {De Ridder}, {Desiati}, {de With}, {DeYoung}, {D{\'\i}az-V{\'e}lez}, {Dunkman}, {Eagan}, {Eberhardt}, {Eisch}, {Ellsworth}, {Euler}, {Evenson}, {Fadiran}, {Fazely}, {Fedynitch}, {Feintzeig}, {Feusels}, {Filimonov}, {Finley}, {Fischer-Wasels}, {Flis}, {Franckowiak}, {Franke}, {Frantzen}, {Fuchs}, {Gaisser}, {Gallagher}, {Gerhardt},
  {Gladstone}, {Gl{\"u}senkamp}, {Goldschmidt}, {Golup}, {Gonzalez}, {Goodman}, {G{\'o}ra}, {Grant}, {Gro{\ss}}, {Gurtner}, {Ha}, {Haj Ismail}, {Hallen}, {Hallgren}, {Halzen}, {Hanson}, {Heereman}, {Heinen}, {Helbing}, {Hellauer}, {Hickford}, {Hill}, {Hoffman}, {Hoffmann}, {Homeier}, {Hoshina}, {Huelsnitz}, {Hulth}, {Hultqvist}, {Hussain}, {Ishihara}, {Jacobi}, {Jacobsen}, {Jagielski}, {Japaridze}, {Jero}, {Jlelati}, {Kaminsky}, {Kappes}, {Karg}, {Karle}, {Kelley}, {Kiryluk}, {Kislat}, {Kl{\"a}s}, {Klein}, {K{\"o}hne}, {Kohnen}, {Kolanoski}, {K{\"o}pke}, {Kopper}, {Kopper}, {Koskinen}, {Kowalski}, {Krasberg}, {Krings}, {Kroll}, {Kunnen}, {Kurahashi}, {Kuwabara}, {Labare}, {Landsman}, {Larson}, {Lesiak-Bzdak}, {Leuermann}, {Leute}, {L{\"u}nemann}, {Madsen}, {Maruyama}, {Mase}, {Matis}, {McNally}, {Meagher}, {Merck}, {M{\'e}sz{\'a}ros}, {Meures}, {Miarecki}, {Middell}, {Milke}, {Miller}, {Mohrmann}, {Montaruli}, {Morse}, {Nahnhauer}, {Naumann}, {Niederhausen}, {Nowicki}, {Nygren}, {Obertacke}, {Odrowski},
  {Olivas}, {Olivo}, {O'Murchadha}, {Paul}, {Pepper}, {P{\'e}rez de los Heros}, {Pfendner}, {Pieloth}, {Pinat}, {Pirk}, {Posselt}, {Price}, {Przybylski}, {R{\"a}del}, {Rameez}, {Rawlins}, {Redl}, {Reimann}, {Resconi}, {Rhode}, {Ribordy}, {Richman}, {Riedel}, {Rodrigues}, {Rott}, {Ruhe}, {Ruzybayev}, {Ryckbosch}, {Saba}, {Salameh}, {Sander}, {Santander}, {Sarkar}, {Schatto}, {Scheel}, {Scheriau}, {Schmidt}, {Schmitz}, {Schoenen}, {Sch{\"o}neberg}, {Sch{\"o}nwald}, {Schukraft}, {Schulte}, {Schulz}, {Seckel}, {Sestayo}, {Seunarine}, {Sheremata}, {Smith}, {Soiron}, {Soldin}, {Spiczak}, {Spiering}, {Stamatikos}, {Stanev}, {Stasik}, {Stezelberger}, {Stokstad}, {St{\"o}{\ss}l}, {Strahler}, {Str{\"o}m}, {Sullivan}, {Taavola}, {Taboada}, {Tamburro}, {Ter-Antonyan}, {Te{\v{s}}i{\'c}}, {Tilav}, {Toale}, {Toscano}, {Usner}, {van der Drift}, {van Eijndhoven}, {Van Overloop}, {van Santen}, {Vehring}, {Voge}, {Vraeghe}, {Walck}, {Waldenmaier}, {Wallraff}, {Wasserman}, {Weaver}, {Wellons}, {Wendt}, {Westerhoff}, {Whitehorn},
  {Wiebe}, {Wiebusch}, {Williams}, {Wissing}, {Wolf}, {Wood}, {Woschnagg}, {Xu}, {Xu}, {Xu}, {Yanez}, {Yodh}, {Yoshida}, {Zarzhitsky}, {Ziemann}, {Zierke}, {Zilles}, \& {Zoll}}]{Aartsen_2013}
{Aartsen}, M.~G., {Abbasi}, R., {Abdou}, Y., {et~al.} 2013, \prl, 111, 021103, \dodoi{10.1103/PhysRevLett.111.021103}

\bibitem[{Abbasi {et~al.}(2021)}]{IceCube:2021oqh}
Abbasi, R., {et~al.} 2021, PoS, ICRC2021, 971, \dodoi{10.22323/1.395.0971}

\bibitem[{{Ansoldi} {et~al.}(2018){Ansoldi}, {Antonelli}, {Arcaro}, {Baack}, {Babi{\'c}}, {Banerjee}, {Bangale}, {Barres de Almeida}, {Barrio}, {Becerra Gonz{\'a}lez}, {Bednarek}, {Bernardini}, {Berse}, {Berti}, {Besenrieder}, {Bhattacharyya}, {Bigongiari}, {Biland}, {Blanch}, {Bonnoli}, {Carosi}, {Ceribella}, {Chatterjee}, {Colak}, {Colin}, {Colombo}, {Contreras}, {Cortina}, {Covino}, {Cumani}, {D'Elia}, {Da Vela}, {Dazzi}, {De Angelis}, {De Lotto}, {Delfino}, {Delgado}, {Di Pierro}, {Dom{\'\i}nguez}, {Dominis Prester}, {Dorner}, {Doro}, {Einecke}, {Elsaesser}, {Fallah Ramazani}, {Fattorini}, {Fern{\'a}ndez-Barral}, {Ferrara}, {Fidalgo}, {Foffano}, {Fonseca}, {Font}, {Fruck}, {Gallozzi}, {Garc{\'\i}a L{\'o}pez}, {Garczarczyk}, {Gaug}, {Giammaria}, {Godinovi{\'c}}, {Guberman}, {Hadasch}, {Hahn}, {Hassan}, {Hayashida}, {Herrera}, {Hoang}, {Hrupec}, {Inoue}, {Ishio}, {Iwamura}, {Konno}, {Kubo}, {Kushida}, {Lamastra}, {Lelas}, {Leone}, {Lindfors}, {Lombardi}, {Longo}, {L{\'o}pez}, {Maggio}, {Majumdar},
  {Makariev}, {Maneva}, {Manganaro}, {Mannheim}, {Maraschi}, {Mariotti}, {Mart{\'\i}nez}, {Masuda}, {Mazin}, {Mielke}, {Minev}, {Miranda}, {Mirzoyan}, {Moralejo}, {Moreno}, {Moretti}, {Neustroev}, {Niedzwiecki}, {Nievas Rosillo}, {Nigro}, {Nilsson}, {Ninci}, {Nishijima}, {Noda}, {Nogu{\'e}s}, {Paiano}, {Palacio}, {Paneque}, {Paoletti}, {Paredes}, {Pedaletti}, {Pe{\~n}il}, {Peresano}, {Persic}, {Pfrang}, {Prada Moroni}, {Prandini}, {Puljak}, {Garcia}, {Rhode}, {Rib{\'o}}, {Rico}, {Righi}, {Rugliancich}, {Saha}, {Saito}, {Satalecka}, {Schweizer}, {Sitarek}, {{\v{S}}nidari{\'c}}, {Sobczynska}, {Stamerra}, {Strzys}, {Suri{\'c}}, {Tavecchio}, {Temnikov}, {Terzi{\'c}}, {Teshima}, {Torres-Alb{\'a}}, {Tsujimoto}, {Vanzo}, {Vazquez Acosta}, {Vovk}, {Ward}, {Will}, {Zari{\'c}}, \& {Cerruti}}]{Ansoldi_2018}
{Ansoldi}, S., {Antonelli}, L.~A., {Arcaro}, C., {et~al.} 2018, \apjl, 863, L10, \dodoi{10.3847/2041-8213/aad083}

\bibitem[{{Atoyan} \& {Dermer}(2001)}]{2001PhRvL..87v1102A}
{Atoyan}, A., \& {Dermer}, C.~D. 2001, \prl, 87, 221102, \dodoi{10.1103/PhysRevLett.87.221102}

\bibitem[{{Baikal-GVD Collaboration} {et~al.}(2022){Baikal-GVD Collaboration}, {Erkenov}, {Kosogorov}, {Kovalev}, {Kovalev}, {Plavin}, {Popkov}, {Pushkarev}, {Semikoz}, {Sotnikova}, \& {Troitsky}}]{Baikal-GVD_2021}
{Baikal-GVD Collaboration}, {Erkenov}, A.~K., {Kosogorov}, N.~A., {et~al.} 2022, arXiv e-prints, arXiv:2210.01650, \dodoi{10.48550/arXiv.2210.01650}

\bibitem[{{Banik} \& {Bhadra}(2019)}]{2019PhRvD..99j3006B}
{Banik}, P., \& {Bhadra}, A. 2019, \prd, 99, 103006, \dodoi{10.1103/PhysRevD.99.103006}

\bibitem[{{Beasley} {et~al.}(2002){Beasley}, {Gordon}, {Peck}, {Petrov}, {MacMillan}, {Fomalont}, \& {Ma}}]{Beasley_2002ApJS}
{Beasley}, A.~J., {Gordon}, D., {Peck}, A.~B., {et~al.} 2002, \apjs, 141, 13, \dodoi{10.1086/339806}

\bibitem[{{Becker Tjus} {et~al.}(2022{\natexlab{a}}){Becker Tjus}, {Jaroschewski}, {Ghorbanietemad}, {Bartos}, {Kun}, \& {Biermann}}]{IceCube_2022}
{Becker Tjus}, J., {Jaroschewski}, I., {Ghorbanietemad}, A., {et~al.} 2022{\natexlab{a}}, arXiv e-prints, arXiv:2210.00202, \dodoi{10.48550/arXiv.2210.00202}

\bibitem[{{Becker Tjus} {et~al.}(2022{\natexlab{b}}){Becker Tjus}, {Jaroschewski}, {Ghorbanietemad}, {Bartos}, {Kun}, \& {Biermann}}]{Becker_2022arXiv}
---. 2022{\natexlab{b}}, arXiv e-prints, arXiv:2210.00202, \dodoi{10.48550/arXiv.2210.00202}

\bibitem[{{Begelman}(1998)}]{Begelman_1998ApJ}
{Begelman}, M.~C. 1998, \apj, 493, 291, \dodoi{10.1086/305119}

\bibitem[{{B{\"o}ttcher} \& {Dermer}(2010)}]{Bottcher_2010ApJ...711..445B}
{B{\"o}ttcher}, M., \& {Dermer}, C.~D. 2010, \apj, 711, 445, \dodoi{10.1088/0004-637X/711/1/445}

\bibitem[{Böttcher {et~al.}(2013)Böttcher, Reimer, Sweeney, \& Prakash}]{Bottcher_2013}
Böttcher, M., Reimer, A., Sweeney, K., \& Prakash, A. 2013, The Astrophysical Journal, 768, 54, \dodoi{10.1088/0004-637x/768/1/54}

\bibitem[{Celotti \& Ghisellini(2008)}]{Celotti_2008}
Celotti, A., \& Ghisellini, G. 2008, Monthly Notices of the Royal Astronomical Society, 385, 283, \dodoi{10.1111/j.1365-2966.2007.12758.x}

\bibitem[{{Cerruti} {et~al.}(2019){Cerruti}, {Zech}, {Boisson}, {Emery}, {Inoue}, \& {Lenain}}]{Cerruti_2019}
{Cerruti}, M., {Zech}, A., {Boisson}, C., {et~al.} 2019, \mnras, 483, L12, \dodoi{10.1093/mnrasl/sly210}

\bibitem[{{Cleary} {et~al.}(2007){Cleary}, {Lawrence}, {Marshall}, {Hao}, \& {Meier}}]{2007ApJ...660..117C}
{Cleary}, K., {Lawrence}, C.~R., {Marshall}, J.~A., {Hao}, L., \& {Meier}, D. 2007, \apj, 660, 117, \dodoi{10.1086/511969}

\bibitem[{{Das} {et~al.}(2022){Das}, {Gupta}, \& {Razzaque}}]{Das_2022}
{Das}, S., {Gupta}, N., \& {Razzaque}, S. 2022, \aap, 668, A146, \dodoi{10.1051/0004-6361/202244653}

\bibitem[{{Deng} {et~al.}(2015){Deng}, {Li}, {Zhang}, \& {Li}}]{Deng_wei_2015}
{Deng}, W., {Li}, H., {Zhang}, B., \& {Li}, S. 2015, \apj, 805, 163, \dodoi{10.1088/0004-637X/805/2/163}

\bibitem[{Foreman-Mackey {et~al.}(2013)Foreman-Mackey, Hogg, Lang, \& Goodman}]{emcee_2013}
Foreman-Mackey, D., Hogg, D.~W., Lang, D., \& Goodman, J. 2013, Publications of the Astronomical Society of the Pacific, 125, 306–312, \dodoi{10.1086/670067}

\bibitem[{{Fromm} {et~al.}(2016){Fromm}, {Perucho}, {Mimica}, \& {Ros}}]{Fromm2016}
{Fromm}, C.~M., {Perucho}, M., {Mimica}, P., \& {Ros}, E. 2016, \aap, 588, A101, \dodoi{10.1051/0004-6361/201527139}

\bibitem[{Förster {et~al.}(2021)Förster, Cabrera-Vives, Castillo-Navarrete, Estévez, Sánchez-Sáez, Arredondo, Bauer, Carrasco-Davis, Catelan, Elorrieta, Eyheramendy, Huijse, Pignata, Reyes, Reyes, Rodríguez-Mancini, Ruz-Mieres, Valenzuela, Álvarez Maldonado, Astorga, Borissova, Clocchiatti, Cicco, Donoso-Oliva, Hernández-García, Graham, Jordán, Kurtev, Mahabal, Maureira, Muñoz-Arancibia, Molina-Ferreiro, Moya, Palma, Pérez-Carrasco, Protopapas, Romero, Sabatini-Gacitua, Sánchez, Martín, Sepúlveda-Cobo, Vera, \& Vergara}]{Forster_2021}
Förster, F., Cabrera-Vives, G., Castillo-Navarrete, E., {et~al.} 2021, The Astronomical Journal, 161, 242, \dodoi{10.3847/1538-3881/abe9bc}

\bibitem[{{Gao} {et~al.}(2019){Gao}, {Fedynitch}, {Winter}, \& {Pohl}}]{2019NatAs...3...88G}
{Gao}, S., {Fedynitch}, A., {Winter}, W., \& {Pohl}, M. 2019, Nature Astronomy, 3, 88, \dodoi{10.1038/s41550-018-0610-1}

\bibitem[{{Gasparyan} {et~al.}(2022){Gasparyan}, {B{\'e}gu{\'e}}, \& {Sahakyan}}]{Gasparyan_2021}
{Gasparyan}, S., {B{\'e}gu{\'e}}, D., \& {Sahakyan}, N. 2022, \mnras, 509, 2102, \dodoi{10.1093/mnras/stab2688}

\bibitem[{{Ghisellini} {et~al.}(2010){Ghisellini}, {Tavecchio}, {Foschini}, {Ghirlanda}, {Maraschi}, \& {Celotti}}]{2010MNRAS.402..497G}
{Ghisellini}, G., {Tavecchio}, F., {Foschini}, L., {et~al.} 2010, \mnras, 402, 497, \dodoi{10.1111/j.1365-2966.2009.15898.x}

\bibitem[{{Giannios} \& {Spruit}(2006)}]{Giannios_2006A&A}
{Giannios}, D., \& {Spruit}, H.~C. 2006, \aap, 450, 887, \dodoi{10.1051/0004-6361:20054107}

\bibitem[{{Giannios} {et~al.}(2009){Giannios}, {Uzdensky}, \& {Begelman}}]{Giannios_2009MNRAG}
{Giannios}, D., {Uzdensky}, D.~A., \& {Begelman}, M.~C. 2009, \mnras, 395, L29, \dodoi{10.1111/j.1745-3933.2009.00635.x}

\bibitem[{Halzen \& Hooper(2002)}]{Halzen_2002}
Halzen, F., \& Hooper, D. 2002, Reports on Progress in Physics, 65, 1025, \dodoi{10.1088/0034-4885/65/7/201}

\bibitem[{Harris \& Krawczynski(2006)}]{Harris_2006}
Harris, D., \& Krawczynski, H. 2006, Annual Review of Astronomy and Astrophysics, 44, 463–506, \dodoi{10.1146/annurev.astro.44.051905.092446}

\bibitem[{{Hayashida} {et~al.}(2012){Hayashida}, {Madejski}, {Nalewajko}, {Sikora}, {Wehrle}, {Ogle}, {Collmar}, {Larsson}, {Fukazawa}, {Itoh}, {Chiang}, {Stawarz}, {Blandford}, {Richards}, {Max-Moerbeck}, {Readhead}, {Buehler}, {Cavazzuti}, {Ciprini}, {Gehrels}, {Reimer}, {Szostek}, {Tanaka}, {Tosti}, {Uchiyama}, {Kawabata}, {Kino}, {Sakimoto}, {Sasada}, {Sato}, {Uemura}, {Yamanaka}, {Greiner}, {Kruehler}, {Rossi}, {Macquart}, {Bock}, {Villata}, {Raiteri}, {Agudo}, {Aller}, {Aller}, {Arkharov}, {Bach}, {Ben{\'\i}tez}, {Berdyugin}, {Blinov}, {Blumenthal}, {B{\"o}ttcher}, {Buemi}, {Carosati}, {Chen}, {Di Paola}, {Dolci}, {Efimova}, {Forn{\'e}}, {G{\'o}mez}, {Gurwell}, {Heidt}, {Hiriart}, {Jordan}, {Jorstad}, {Joshi}, {Kimeridze}, {Konstantinova}, {Kopatskaya}, {Koptelova}, {Kurtanidze}, {L{\"a}hteenm{\"a}ki}, {Lamerato}, {Larionov}, {Larionova}, {Larionova}, {Leto}, {Lindfors}, {Marscher}, {McHardy}, {Molina}, {Morozova}, {Nikolashvili}, {Nilsson}, {Reinthal}, {Roustazadeh}, {Sakamoto}, {Sigua},
  {Sillanp{\"a}{\"a}}, {Takalo}, {Tammi}, {Taylor}, {Tornikoski}, {Trigilio}, {Troitsky}, \& {Umana}}]{2012ApJ...754..114H}
{Hayashida}, M., {Madejski}, G.~M., {Nalewajko}, K., {et~al.} 2012, \apj, 754, 114, \dodoi{10.1088/0004-637X/754/2/114}

\bibitem[{{Heckman} \& {Best}(2014)}]{Heckman_2014ARA&A}
{Heckman}, T.~M., \& {Best}, P.~N. 2014, \araa, 52, 589, \dodoi{10.1146/annurev-astro-081913-035722}

\bibitem[{{IceCube Collaboration}(2013)}]{IceCube_2013}
{IceCube Collaboration}. 2013, Science, 342, 1242856, \dodoi{10.1126/science.1242856}

\bibitem[{{IceCube Collaboration} {et~al.}(2018{\natexlab{a}}){IceCube Collaboration}, {Aartsen}, {Ackermann}, {Adams}, {Aguilar}, {Ahlers}, {Ahrens}, {Al Samarai}, {Altmann}, {Andeen}, {Anderson}, {Ansseau}, {Anton}, {Arg{\"u}elles}, {Auffenberg}, {Axani}, {Bagherpour}, {Bai}, {Barron}, {Barwick}, {Baum}, {Bay}, {Beatty}, {Becker Tjus}, {Becker}, {BenZvi}, {Berley}, {Bernardini}, {Besson}, {Binder}, {Bindig}, {Blaufuss}, {Blot}, {Bohm}, {B{\"o}rner}, {Bos}, {B{\"o}ser}, {Botner}, {Bourbeau}, {Bourbeau}, {Bradascio}, {Braun}, {Brenzke}, {Bretz}, {Bron}, {Brostean-Kaiser}, {Burgman}, {Busse}, {Carver}, {Cheung}, {Chirkin}, {Christov}, {Clark}, {Classen}, {Coenders}, {Collin}, {Conrad}, {Coppin}, {Correa}, {Cowen}, {Cross}, {Dave}, {Day}, {de Andr{\'e}}, {De Clercq}, {DeLaunay}, {Dembinski}, {De Ridder}, {Desiati}, {de Vries}, {de Wasseige}, {de With}, {DeYoung}, {D{\'\i}az-V{\'e}lez}, {di Lorenzo}, {Dujmovic}, {Dumm}, {Dunkman}, {Dvorak}, {Eberhardt}, {Ehrhardt}, {Eichmann}, {Eller}, {Evenson}, {Fahey},
  {Fazely}, {Felde}, {Filimonov}, {Finley}, {Flis}, {Franckowiak}, {Friedman}, {Fritz}, {Gaisser}, {Gallagher}, {Gerhardt}, {Ghorbani}, {Glauch}, {Gl{\"u}senkamp}, {Goldschmidt}, {Gonzalez}, {Grant}, {Griffith}, {Haack}, {Hallgren}, {Halzen}, {Hanson}, {Hebecker}, {Heereman}, {Helbing}, {Hellauer}, {Hickford}, {Hignight}, {Hill}, {Hoffman}, {Hoffmann}, {Hoinka}, {Hokanson-Fasig}, {Hoshina}, {Huang}, {Huber}, {Hultqvist}, {H{\"u}nnefeld}, {Hussain}, {In}, {Iovine}, {Ishihara}, {Jacobi}, {Japaridze}, {Jeong}, {Jero}, {Jones}, {Kalaczynski}, {Kang}, {Kappes}, {Kappesser}, {Karg}, {Karle}, {Katz}, {Kauer}, {Keivani}, {Kelley}, {Kheirandish}, {Kim}, {Kim}, {Kintscher}, {Kiryluk}, {Kittler}, {Klein}, {Koirala}, {Kolanoski}, {K{\"o}pke}, {Kopper}, {Kopper}, {Koschinsky}, {Koskinen}, {Kowalski}, {Krings}, {Kroll}, {Kr{\"u}ckl}, {Kunwar}, {Kurahashi}, {Kuwabara}, {Kyriacou}, {Labare}, {Lanfranchi}, {Larson}, {Lauber}, {Leonard}, {Lesiak-Bzdak}, {Leuermann}, {Liu}, {Lozano Mariscal}, {Lu}, {L{\"u}nemann}, {Luszczak},
  {Madsen}, {Maggi}, {Mahn}, {Mancina}, {Maruyama}, {Mase}, {Maunu}, {Meagher}, {Medici}, {Meier}, {Menne}, {Merino}, {Meures}, {Miarecki}, {Micallef}, {Moment{\'e}}, {Montaruli}, {Moore}, {Morse}, {Moulai}, {Nahnhauer}, {Nakarmi}, {Naumann}, {Neer}, {Niederhausen}, {Nowicki}, {Nygren}, {Obertacke Pollmann}, {Olivas}, {O'Murchadha}, {O'Sullivan}, {Palczewski}, {Pandya}, {Pankova}, {Peiffer}, {Pepper}, {P{\'e}rez de los Heros}, {Pieloth}, {Pinat}, {Plum}, {Price}, {Przybylski}, {Raab}, {R{\"a}del}, {Rameez}, {Rauch}, {Rawlins}, {Rea}, {Reimann}, {Relethford}, {Relich}, {Resconi}, {Rhode}, {Richman}, {Robertson}, {Rongen}, {Rott}, {Ruhe}, {Ryckbosch}, {Rysewyk}, {Safa}, {S{\"a}lzer}, {Sanchez Herrera}, {Sandrock}, {Sandroos}, {Santander}, {Sarkar}, {Sarkar}, {Satalecka}, {Schlunder}, {Schmidt}, {Schneider}, {Schoenen}, {Sch{\"o}neberg}, {Schumacher}, {Sclafani}, {Seckel}, {Seunarine}, {Soedingrekso}, {Soldin}, {Song}, {Spiczak}, {Spiering}, {Stachurska}, {Stamatikos}, {Stanev}, {Stasik}, {Stein}, {Stettner},
  {Steuer}, {Stezelberger}, {Stokstad}, {St{\"o}{\ss}l}, {Strotjohann}, {Stuttard}, {Sullivan}, {Sutherland}, {Taboada}, {Tatar}, {Tenholt}, {Ter-Antonyan}, {Terliuk}, {Tilav}, {Toale}, {Tobin}, {Toennis}, {Toscano}, {Tosi}, {Tselengidou}, {Tung}, {Turcati}, {Turley}, {Ty}, {Unger}, {Usner}, {Vandenbroucke}, {Van Driessche}, {van Eijk}, {van Eijndhoven}, {Vanheule}, {van Santen}, {Vogel}, {Vraeghe}, {Walck}, {Wallace}, {Wallraff}, {Wandler}, {Wandkowsky}, {Waza}, {Weaver}, {Weiss}, {Wendt}, {Werthebach}, {Westerhoff}, {Whelan}, {Whitehorn}, {Wiebe}, {Wiebusch}, {Wille}, {Williams}, {Wills}, {Wolf}, {Wood}, {Wood}, {Woschnagg}, {Xu}, {Xu}, {Xu}, {Yanez}, {Yodh}, {Yoshida}, {Yuan}, {Fermi-LAT Collaboration}, {Abdollahi}, {Ajello}, {Angioni}, {Baldini}, {Ballet}, {Barbiellini}, {Bastieri}, {Bechtol}, {Bellazzini}, {Berenji}, {Bissaldi}, {Blandford}, {Bonino}, {Bottacini}, {Bregeon}, {Bruel}, {Buehler}, {Burnett}, {Burns}, {Buson}, {Cameron}, {Caputo}, {Caraveo}, {Cavazzuti}, {Charles}, {Chen}, {Cheung},
  {Chiang}, {Chiaro}, {Ciprini}, {Cohen-Tanugi}, {Conrad}, {Costantin}, {Cutini}, {D'Ammando}, {de Palma}, {Digel}, {Di Lalla}, {Di Mauro}, {Di Venere}, {Dom{\'\i}nguez}, {Favuzzi}, {Franckowiak}, {Fukazawa}, {Funk}, {Fusco}, {Gargano}, {Gasparrini}, {Giglietto}, {Giomi}, {Giommi}, {Giordano}, {Giroletti}, {Glanzman}, {Green}, {Grenier}, {Grondin}, {Guiriec}, {Harding}, {Hayashida}, {Hays}, {Hewitt}, {Horan}, {J{\'o}hannesson}, {Kadler}, {Kensei}, {Kocevski}, {Krauss}, {Kreter}, {Kuss}, {La Mura}, {Larsson}, {Latronico}, {Lemoine-Goumard}, {Li}, {Longo}, {Loparco}, {Lovellette}, {Lubrano}, {Magill}, {Maldera}, {Malyshev}, {Manfreda}, {Mazziotta}, {McEnery}, {Meyer}, {Michelson}, {Mizuno}, {Monzani}, {Morselli}, {Moskalenko}, {Negro}, {Nuss}, {Ojha}, {Omodei}, {Orienti}, {Orlando}, {Palatiello}, {Paliya}, {Perkins}, {Persic}, {Pesce-Rollins}, {Piron}, {Porter}, {Principe}, {Rain{\`o}}, {Rando}, {Rani}, {Razzano}, {Razzaque}, {Reimer}, {Reimer}, {Renault-Tinacci}, {Ritz}, {Rochester}, {Saz Parkinson},
  {Sgr{\`o}}, {Siskind}, {Spandre}, {Spinelli}, {Suson}, {Tajima}, {Takahashi}, {Tanaka}, {Thayer}, {Thompson}, {Tibaldo}, {Torres}, {Torresi}, {Tosti}, {Troja}, {Valverde}, {Vianello}, {Vogel}, {Wood}, {Wood}, {Zaharijas}, {MAGIC Collaboration}, {Ahnen}, {Ansoldi}, {Antonelli}, {Arcaro}, {Baack}, {Babi{\'c}}, {Banerjee}, {Bangale}, {Barres de Almeida}, {Barrio}, {Becerra Gonz{\'a}lez}, {Bednarek}, {Bernardini}, {Berti}, {Bhattacharyya}, {Biland}, {Blanch}, {Bonnoli}, {Carosi}, {Carosi}, {Ceribella}, {Chatterjee}, {Colak}, {Colin}, {Colombo}, {Contreras}, {Cortina}, {Covino}, {Cumani}, {Da Vela}, {Dazzi}, {De Angelis}, {De Lotto}, {Delfino}, {Delgado}, {Di Pierro}, {Dom{\'\i}nguez}, {Dominis Prester}, {Dorner}, {Doro}, {Einecke}, {Elsaesser}, {Fallah Ramazani}, {Fern{\'a}ndez-Barral}, {Fidalgo}, {Foffano}, {Pfrang}, {Fonseca}, {Font}, {Franceschini}, {Fruck}, {Galindo}, {Gallozzi}, {Garc{\'\i}a L{\'o}pez}, {Garczarczyk}, {Gaug}, {Giammaria}, {Godinovi{\'c}}, {Gora}, {Guberman}, {Hadasch}, {Hahn}, {Hassan},
  {Hayashida}, {Herrera}, {Hose}, {Hrupec}, {Inoue}, {Ishio}, {Konno}, {Kubo}, {Kushida}, {Lelas}, {Lindfors}, {Lombardi}, {Longo}, {L{\'o}pez}, {Maggio}, {Majumdar}, {Makariev}, {Maneva}, {Manganaro}, {Mannheim}, {Maraschi}, {Mariotti}, {Mart{\'\i}nez}, {Masuda}, {Mazin}, {Minev}, {M}, {Mirzoyan}, {Moralejo}, {Moreno}, {Moretti}, {Nagayoshi}, {Neustroev}, {Niedzwiecki}, {Nievas Rosillo}, {Nigro}, {Nilsson}, {Ninci}, {Nishijima}, {Noda}, {Nogu{\'e}s}, {Paiano}, {Palacio}, {Paneque}, {Paoletti}, {Paredes}, {Pedaletti}, {Peresano}, {Persic}, {Prada Moroni}, {Prandini}, {Puljak}, {Rodriguez Garcia}, {Reichardt}, {Rhode}, {Rib{\'o}}, {Rico}, {Righi}, {Rugliancich}, {Saito}, {Satalecka}, {Schweizer}, {Sitarek}, {{\v{S}}nidaric {\textasciiacute}}, {Sobczynska}, {Stamerra}, {Strzys}, {Suri{\'c}}, {Takahashi}, {Tavecchio}, {Temnikov}, {Terzi{\'c}}, {Teshima}, {Torres-Alb{\`a}}, {Treves}, {Tsujimoto}, {Vanzo}, {Vazquez Acosta}, {Vovk}, {Ward}, {Will}, {S}, {Zaric {\textasciiacute}}, {AGILE Team}, {Lucarelli},
  {Tavani}, {Piano}, {Donnarumma}, {Pittori}, {Verrecchia}, {Barbiellini}, {Bulgarelli}, {Caraveo}, {Cattaneo}, {Colafrancesco}, {Costa}, {Di Cocco}, {Ferrari}, {Gianotti}, {Giuliani}, {Lipari}, {Mereghetti}, {Morselli}, {Pacciani}, {Paoletti}, {Parmiggiani}, {Pellizzoni}, {Picozza}, {Pilia}, {Rappoldi}, {Trois}, {Vercellone}, {Vittorini}, {ASAS-SN Team}, {Stanek}, {Franckowiak}, {Kochanek}, {Beacom}, {Thompson}, {Holoien}, {Dong}, {Prieto}, {Shappee}, {Holmbo}, {HAWC Collaboration}, {Abeysekara}, {Albert}, {Alfaro}, {Alvarez}, {Arceo}, {Arteaga-Vel{\'a}zquez}, {Avila Rojas}, {Ayala Solares}, {Becerril}, {Belmont-Moreno}, {Bernal}, {Caballero-Mora}, {Capistr{\'a}n}, {Carrami{\~n}ana}, {Casanova}, {Castillo}, {Cotti}, {Cotzomi}, {Couti{\~n}o de Le{\'o}n}, {De Le{\'o}n}, {De la Fuente}, {Diaz Hernandez}, {Dichiara}, {Dingus}, {DuVernois}, {D{\'\i}az-V{\'e}lez}, {Ellsworth}, {Engel}, {Fiorino}, {Fleischhack}, {Fraija}, {Garc{\'\i}a-Gonz{\'a}lez}, {Garfias}, {Gonz{\'a}lez Mu{\~n}oz}, {Gonz{\'a}lez}, {Goodman},
  {Hampel-Arias}, {Harding}, {Hernandez}, {Hona}, {Hueyotl-Zahuantitla}, {Hui}, {H{\"u}ntemeyer}, {Iriarte}, {Jardin-Blicq}, {Joshi}, {Kaufmann}, {Kunde}, {Lara}, {Lauer}, {Lee}, {Lennarz}, {Le{\'o}n Vargas}, {Linnemann}, {Longinotti}, {Luis-Raya}, {Luna-Garc{\'\i}a}, {Malone}, {Marinelli}, {Martinez}, {Martinez-Castellanos}, {Mart{\'\i}nez-Castro}, {Mart{\'\i}nez-Huerta}, {Matthews}, {Miranda-Romagnoli}, {Moreno}, {Mostaf{\'a}}, {Nayerhoda}, {Nellen}, {Newbold}, {Nisa}, {Noriega-Papaqui}, {Pelayo}, {Pretz}, {P{\'e}rez-P{\'e}rez}, {Ren}, {Rho}, {Rivi{\`e}re}, {Rosa-Gonz{\'a}lez}, {Rosenberg}, {Ruiz-Velasco}, {Ruiz-Velasco}, {Salesa Greus}, {Sandoval}, {Schneider}, {Schoorlemmer}, {Sinnis}, {Smith}, {Springer}, {Surajbali}, {Tibolla}, {Tollefson}, {Torres}, {Villase{\~n}or}, {Weisgarber}, {Werner}, {Yapici}, {Gaurang}, {Zepeda}, {Zhou}, {{\'A}lvarez}, {H.~E.~S.~S. Collaboration}, {Abdalla}, {Ang{\"u}ner}, {Armand}, {Backes}, {Becherini}, {Berge}, {B{\"o}ttcher}, {Boisson}, {Bolmont}, {Bonnefoy}, {Bordas},
  {Brun}, {B{\"u}chele}, {Bulik}, {Caroff}, {Carosi}, {Casanova}, {Cerruti}, {Chakraborty}, {Chandra}, {Chen}, {Colafrancesco}, {Davids}, {Deil}, {Devin}, {Djannati-Ata{\"\i}}, {Egberts}, {Emery}, {Eschbach}, {Fiasson}, {Fontaine}, {Funk}, {F{\"u}{\ss}ling}, {Gallant}, {Gat{\'e}}, {Giavitto}, {Glawion}, {Glicenstein}, {Gottschall}, {Grondin}, {Haupt}, {Henri}, {Hinton}, {Hoischen}, {Holch}, {Huber}, {Jamrozy}, {Jankowsky}, {Jankowsky}, {Jouvin}, {Jung-Richardt}, {Kerszberg}, {Kh{\'e}lifi}, {King}, {Klepser}, {Kluz {\textasciiacute}niak}, {Komin}, {Kraus}, {Lefaucheur}, {Lemi{\`e}re}, {Lemoine-Goumard}, {Lenain}, {Leser}, {Lohse}, {L{\'o}pez-Coto}, {Lorentz}, {Lypova}, {Marandon}, {Guillem Mart{\'\i}-Devesa}, {Maurin}, {Mitchell}, {Moderski}, {Mohamed}, {Mohrmann}, {Moulin}, {Murach}, {de Naurois}, {Niederwanger}, {Niemiec}, {Oakes}, {O'Brien}, {Ohm}, {Ostrowski}, {Oya}, {Panter}, {Parsons}, {Perennes}, {Piel}, {Pita}, {Poireau}, {Priyana Noel}, {Prokoph}, {P{\"u}hlhofer}, {Quirrenbach}, {Raab}, {Rauth},
  {Renaud}, {Rieger}, {Rinchiuso}, {Romoli}, {Rowell}, {Rudak}, {Sasaki}, {Sanchez}, {Schlickeiser}, {Sch{\"u}ssler}, {Schulz}, {Schwanke}, {Seglar-Arroyo}, {Shafi}, {Simoni}, {Sol}, {Stegmann}, {Steppa}, {Tavernier}, {Taylor}, {Tiziani}, {Trichard}, {Tsirou}, {van Eldik}, {van Rensburg}, {van Soelen}, {Veh}, {Vincent}, {Voisin}, {Wagner}, {Wagner}, {Wierzcholska}, {Zanin}, {Zdziarski}, {Zech}, {Ziegler}, {Zorn}, {{\.Z}ywucka}, {INTEGRAL Team}, {Savchenko}, {Ferrigno}, {Bazzano}, {Diehl}, {Kuulkers}, {Laurent}, {Mereghetti}, {Natalucci}, {Panessa}, {Rodi}, {Ubertini}, {Kanata}, Teams, {Morokuma}, {Ohta}, {Tanaka}, {Mori}, {Yamanaka}, {Kawabata}, {Utsumi}, {Nakaoka}, {Kawabata}, {Nagashima}, {Yoshida}, {Matsuoka}, {Itoh}, {Kapteyn Team}, {Keel}, {Liverpool Telescope Team}, {Copperwheat}, {Steele}, {Swift/NuSTAR Team}, {Cenko}, {Cowen}, {DeLaunay}, {Evans}, {Fox}, {Keivani}, {Kennea}, {Marshall}, {Osborne}, {Santander}, {Tohuvavohu}, {Turley}, {VERITAS Collaboration}, {Abeysekara}, {Archer}, {Benbow}, {Bird},
  {Brill}, {Brose}, {Buchovecky}, {Buckley}, {Bugaev}, {Christiansen}, {Connolly}, {Cui}, {Daniel}, {Errando}, {Falcone}, {Feng}, {Finley}, {Fortson}, {Furniss}, {Gueta}, {H{\"u}tten}, {Hervet}, {Hughes}, {Humensky}, {Johnson}, {Kaaret}, {Kar}, {Kelley-Hoskins}, {Kertzman}, {Kieda}, {Krause}, {Krennrich}, {Kumar}, {Lang}, {Lin}, {Maier}, {McArthur}, {Moriarty}, {Mukherjee}, {Nieto}, {O'Brien}, {Ong}, {Otte}, {Park}, {Petrashyk}, {Pohl}, {Popkow}, {Pueschel}, {Quinn}, {Ragan}, {Reynolds}, {Richards}, {Roache}, {Rulten}, {Sadeh}, {Santander}, {Scott}, {Sembroski}, {Shahinyan}, {Sushch}, {Tr{\'e}panier}, {Tyler}, {Vassiliev}, {Wakely}, {Weinstein}, {Wells}, {Wilcox}, {Wilhelm}, {Williams}, {Zitzer}, {VLA/B Team}, {Tetarenko}, {Kimball}, {Miller-Jones}, \& {Sivakoff}}]{2018Sci...361.1378I}
{IceCube Collaboration}, {Aartsen}, M.~G., {Ackermann}, M., {et~al.} 2018{\natexlab{a}}, Science, 361, eaat1378, \dodoi{10.1126/science.aat1378}

\bibitem[{{IceCube Collaboration} {et~al.}(2018{\natexlab{b}}){IceCube Collaboration}, {Aartsen}, {Ackermann}, {Adams}, {Aguilar}, {Ahlers}, {Ahrens}, {Samarai}, {Altmann}, {Andeen}, {Anderson}, {Ansseau}, {Anton}, {Arg{\"u}elles}, {Arsioli}, {Auffenberg}, {Axani}, {Bagherpour}, {Bai}, {Barron}, {Barwick}, {Baum}, {Bay}, {Beatty}, {Becker Tjus}, {Becker}, {BenZvi}, {Berley}, {Bernardini}, {Besson}, {Binder}, {Bindig}, {Blaufuss}, {Blot}, {Bohm}, {B{\"o}rner}, {Bos}, {B{\"o}ser}, {Botner}, {Bourbeau}, {Bourbeau}, {Bradascio}, {Braun}, {Brenzke}, {Bretz}, {Bron}, {Brostean-Kaiser}, {Burgman}, {Busse}, {Carver}, {Cheung}, {Chirkin}, {Christov}, {Clark}, {Classen}, {Coenders}, {Collin}, {Conrad}, {Coppin}, {Correa}, {Cowen}, {Cross}, {Dave}, {Day}, {de Andr{\'e}}, {De Clercq}, {DeLaunay}, {Dembinski}, {DeRidder}, {Desiati}, {de Vries}, {de Wasseige}, {de With}, {DeYoung}, {D{\'\i}az-V{\'e}lez}, {di Lorenzo}, {Dujmovic}, {Dumm}, {Dunkman}, {Dvorak}, {Eberhardt}, {Ehrhardt}, {Eichmann}, {Eller}, {Evenson}, {Fahey},
  {Fazely}, {Felde}, {Filimonov}, {Finley}, {Flis}, {Franckowiak}, {Friedman}, {Fritz}, {Gaisser}, {Gallagher}, {Gerhardt}, {Ghorbani}, {Giommi}, {Glauch}, {Gl{\"u}senkamp}, {Goldschmidt}, {Gonzalez}, {Grant}, {Griffith}, {Haack}, {Hallgren}, {Halzen}, {Hanson}, {Hebecker}, {Heereman}, {Helbing}, {Hellauer}, {Hickford}, {Hignight}, {Hill}, {Hoffman}, {Hoffmann}, {Hoinka}, {Hokanson-Fasig}, {Hoshina}, {Huang}, {Huber}, {Hultqvist}, {H{\"u}nnefeld}, {Hussain}, {In}, {Iovine}, {Ishihara}, {Jacobi}, {Japaridze}, {Jeong}, {Jero}, {Jones}, {Kalaczynski}, {Kang}, {Kappes}, {Kappesser}, {Karg}, {Karle}, {Katz}, {Kauer}, {Keivani}, {Kelley}, {Kheirandish}, {Kim}, {Kim}, {Kintscher}, {Kiryluk}, {Kittler}, {Klein}, {Koirala}, {Kolanoski}, {K{\"o}pke}, {Kopper}, {Kopper}, {Koschinsky}, {Koskinen}, {Kowalski}, {Krammer}, {Krings}, {Kroll}, {Kr{\"u}ckl}, {Kunwar}, {Kurahashi}, {Kuwabara}, {Kyriacou}, {Labare}, {Lanfranchi}, {Larson}, {Lauber}, {Leonard}, {Lesiak-Bzdak}, {Leuermann}, {Liu}, {Lozano Mariscal}, {Lu},
  {L{\"u}nemann}, {Luszczak}, {Madsen}, {Maggi}, {Mahn}, {Mancina}, {Maruyama}, {Mase}, {Maunu}, {Meagher}, {Medici}, {Meier}, {Menne}, {Merino}, {Meures}, {Miarecki}, {Micallef}, {Moment{\'e}}, {Montaruli}, {Moore}, {Morse}, {Moulai}, {Nahnhauer}, {Nakarmi}, {Naumann}, {Neer}, {Niederhausen}, {Nowicki}, {Nygren}, {Obertacke Pollmann}, {Olivas}, {O'Murchadha}, {O'Sullivan}, {Padovani}, {Palczewski}, {Pandya}, {Pankova}, {Peiffer}, {Pepper}, {P{\'e}rez de los Heros}, {Pieloth}, {Pinat}, {Plum}, {Price}, {Przybylski}, {Raab}, {R{\"a}del}, {Rameez}, {Rawlins}, {Rea}, {Reimann}, {Relethford}, {Relich}, {Resconi}, {Rhode}, {Richman}, {Robertson}, {Rongen}, {Rott}, {Ruhe}, {Ryckbosch}, {Rysewyk}, {Safa}, {Sahakyan}, {S{\"a}lzer}, {Sanchez Herrera}, {Sandrock}, {Sandroos}, {Santander}, {Sarkar}, {Sarkar}, {Satalecka}, {Schlunder}, {Schmidt}, {Schneider}, {Schoenen}, {Sch{\"o}neberg}, {Schumacher}, {Sclafani}, {Seckel}, {Seunarine}, {Soedingrekso}, {Soldin}, {Song}, {Spiczak}, {Spiering}, {Stachurska}, {Stamatikos},
  {Stanev}, {Stasik}, {Stettner}, {Steuer}, {Stezelberger}, {Stokstad}, {St{\"o}{\ss}l}, {Strotjohann}, {Stuttard}, {Sullivan}, {Sutherland}, {Taboada}, {Tatar}, {Tenholt}, {Ter-Antonyan}, {Terliuk}, {Tilav}, {Toale}, {Tobin}, {Toennis}, {Toscano}, {Tosi}, {Tselengidou}, {Tung}, {Turcati}, {Turley}, {Ty}, {Unger}, {Usner}, {Vandenbroucke}, {Van Driessche}, {van Eijk}, {van Eijndhoven}, {Vanheule}, {van Santen}, {Vogel}, {Vraeghe}, {Walck}, {Wallace}, {Wallraff}, {Wandler}, {Wandkowsky}, {Waza}, {Weaver}, {Weiss}, {Wendt}, {Werthebach}, {Westerhoff}, {Whelan}, {Whitehorn}, {Wiebe}, {Wiebusch}, {Wille}, {Williams}, {Wills}, {Wolf}, {Wood}, {Wood}, {Woschnagg}, {Xu}, {Xu}, {Xu}, {Yanez}, {Yodh}, {Yoshida}, \& {Yuan}}]{2018Sci...361..147I}
---. 2018{\natexlab{b}}, Science, 361, 147, \dodoi{10.1126/science.aat2890}

\bibitem[{{Joshi} \& {B{\"o}ttcher}(2011)}]{Joshi_2011ApJ}
{Joshi}, M., \& {B{\"o}ttcher}, M. 2011, \apj, 727, 21, \dodoi{10.1088/0004-637X/727/1/21}

\bibitem[{{Kamraj} {et~al.}(2018){Kamraj}, {Harrison}, {Balokovi{\'c}}, {Lohfink}, \& {Brightman}}]{Kamraj_2018ApJ...866..124K}
{Kamraj}, N., {Harrison}, F.~A., {Balokovi{\'c}}, M., {Lohfink}, A., \& {Brightman}, M. 2018, \apj, 866, 124, \dodoi{10.3847/1538-4357/aadd0d}

\bibitem[{{Keivani} {et~al.}(2018){Keivani}, {Murase}, {Petropoulou}, {Fox}, {Cenko}, {Chaty}, {Coleiro}, {DeLaunay}, {Dimitrakoudis}, {Evans}, {Kennea}, {Marshall}, {Mastichiadis}, {Osborne}, {Santander}, {Tohuvavohu}, \& {Turley}}]{2018ApJ...864...84K}
{Keivani}, A., {Murase}, K., {Petropoulou}, M., {et~al.} 2018, \apj, 864, 84, \dodoi{10.3847/1538-4357/aad59a}

\bibitem[{{Kelner} \& {Aharonian}(2008)}]{2008PhRvD..78c4013K}
{Kelner}, S.~R., \& {Aharonian}, F.~A. 2008, \prd, 78, 034013, \dodoi{10.1103/PhysRevD.78.034013}

\bibitem[{Liu {et~al.}(2019)Liu, Wang, Xue, Taylor, Wang, Li, \& Yan}]{Liu_2019}
Liu, R.-Y., Wang, K., Xue, R., {et~al.} 2019, Physical Review D, 99, \dodoi{10.1103/physrevd.99.063008}

\bibitem[{{Liu} {et~al.}(2023){Liu}, {Xue}, {Wang}, {Tan}, \& {B{\"o}ttcher}}]{Ruo_Yu_Liu_2023MNRAS.526.5054L}
{Liu}, R.-Y., {Xue}, R., {Wang}, Z.-R., {Tan}, H.-B., \& {B{\"o}ttcher}, M. 2023, \mnras, 526, 5054, \dodoi{10.1093/mnras/stad2911}

\bibitem[{{Mannheim}(1993)}]{1993A&A...269...67M}
{Mannheim}, K. 1993, \aap, 269, 67.
\newblock \doarXiv{astro-ph/9302006}

\bibitem[{{Mannheim}(1995)}]{1995APh.....3..295M}
---. 1995, Astroparticle Physics, 3, 295, \dodoi{10.1016/0927-6505(94)00044-4}

\bibitem[{{Massaro} {et~al.}(2004){Massaro}, {Perri}, {Giommi}, \& {Nesci}}]{Massaro_2004A&A}
{Massaro}, E., {Perri}, M., {Giommi}, P., \& {Nesci}, R. 2004, \aap, 413, 489, \dodoi{10.1051/0004-6361:20031558}

\bibitem[{{McKinney} \& {Blandford}(2009)}]{McKinney_2009MNRA}
{McKinney}, J.~C., \& {Blandford}, R.~D. 2009, \mnras, 394, L126, \dodoi{10.1111/j.1745-3933.2009.00625.x}

\bibitem[{Meyer {et~al.}(2019)Meyer, Scargle, \& Blandford}]{Meyer_2019}
Meyer, M., Scargle, J.~D., \& Blandford, R.~D. 2019, The Astrophysical Journal, 877, 39, \dodoi{10.3847/1538-4357/ab1651}

\bibitem[{{Moderski} {et~al.}(2005){Moderski}, {Sikora}, {Coppi}, \& {Aharonian}}]{2005MNRAS.363..954M}
{Moderski}, R., {Sikora}, M., {Coppi}, P.~S., \& {Aharonian}, F. 2005, \mnras, 363, 954, \dodoi{10.1111/j.1365-2966.2005.09494.x}

\bibitem[{{Murase} {et~al.}(2014){Murase}, {Inoue}, \& {Dermer}}]{Murase_2014PhRvD..90b3007M}
{Murase}, K., {Inoue}, Y., \& {Dermer}, C.~D. 2014, \prd, 90, 023007, \dodoi{10.1103/PhysRevD.90.023007}

\bibitem[{{Murase} {et~al.}(2018){Murase}, {Oikonomou}, \& {Petropoulou}}]{2018ApJ...865..124M}
{Murase}, K., {Oikonomou}, F., \& {Petropoulou}, M. 2018, \apj, 865, 124, \dodoi{10.3847/1538-4357/aada00}

\bibitem[{{Murase} \& {Stecker}(2022)}]{Murase_2022arXiv220203381M}
{Murase}, K., \& {Stecker}, F.~W. 2022, arXiv e-prints, arXiv:2202.03381, \dodoi{10.48550/arXiv.2202.03381}

\bibitem[{{Nalewajko}(2012)}]{Nalewajko2012}
{Nalewajko}, K. 2012, \mnras, 420, L48, \dodoi{10.1111/j.1745-3933.2011.01193.x}

\bibitem[{Nalewajko \& Begelman(2012)}]{Nalewajko_10.1111}
Nalewajko, K., \& Begelman, M.~C. 2012, Monthly Notices of the Royal Astronomical Society, 427, 2480, \dodoi{10.1111/j.1365-2966.2012.22117.x}

\bibitem[{{O'Sullivan} \& {Gabuzda}(2009)}]{O'Sullivan_2009MN}
{O'Sullivan}, S.~P., \& {Gabuzda}, D.~C. 2009, \mnras, 400, 26, \dodoi{10.1111/j.1365-2966.2009.15428.x}

\bibitem[{{Padovani} {et~al.}(2018){Padovani}, {Giommi}, {Resconi}, {Glauch}, {Arsioli}, {Sahakyan}, \& {Huber}}]{2018MNRAS.480..192P}
{Padovani}, P., {Giommi}, P., {Resconi}, E., {et~al.} 2018, \mnras, 480, 192, \dodoi{10.1093/mnras/sty1852}

\bibitem[{{Padovani} {et~al.}(2019){Padovani}, {Oikonomou}, {Petropoulou}, {Giommi}, \& {Resconi}}]{Padovani_2019MNRA}
{Padovani}, P., {Oikonomou}, F., {Petropoulou}, M., {Giommi}, P., \& {Resconi}, E. 2019, \mnras, 484, L104, \dodoi{10.1093/mnrasl/slz011}

\bibitem[{Paiano {et~al.}(2018)Paiano, Falomo, Treves, \& Scarpa}]{Paiano_2018}
Paiano, S., Falomo, R., Treves, A., \& Scarpa, R. 2018, The Astrophysical Journal Letters, 854, L32, \dodoi{10.3847/2041-8213/aaad5e}

\bibitem[{{Petropoulou} {et~al.}(2016){Petropoulou}, {Giannios}, \& {Sironi}}]{Petropoulou_2016MNRA}
{Petropoulou}, M., {Giannios}, D., \& {Sironi}, L. 2016, \mnras, 462, 3325, \dodoi{10.1093/mnras/stw1832}

\bibitem[{{Reimer} {et~al.}(2019){Reimer}, {B{\"o}ttcher}, \& {Buson}}]{2019ApJ...881...46R}
{Reimer}, A., {B{\"o}ttcher}, M., \& {Buson}, S. 2019, \apj, 881, 46, \dodoi{10.3847/1538-4357/ab2bff}

\bibitem[{{Ricci} {et~al.}(2018){Ricci}, {Ho}, {Fabian}, {Trakhtenbrot}, {Koss}, {Ueda}, {Lohfink}, {Shimizu}, {Bauer}, {Mushotzky}, {Schawinski}, {Paltani}, {Lamperti}, {Treister}, \& {Oh}}]{Ricci_2018MNR}
{Ricci}, C., {Ho}, L.~C., {Fabian}, A.~C., {et~al.} 2018, \mnras, 480, 1819, \dodoi{10.1093/mnras/sty1879}

\bibitem[{{Rieger} {et~al.}(2007){Rieger}, {Bosch-Ramon}, \& {Duffy}}]{Rieger_2007Ap&SS}
{Rieger}, F.~M., {Bosch-Ramon}, V., \& {Duffy}, P. 2007, \apss, 309, 119, \dodoi{10.1007/s10509-007-9466-z}

\bibitem[{{Righi} {et~al.}(2019){Righi}, {Tavecchio}, \& {Inoue}}]{Righi_2019MNR}
{Righi}, C., {Tavecchio}, F., \& {Inoue}, S. 2019, \mnras, 483, L127, \dodoi{10.1093/mnrasl/sly231}

\bibitem[{{Rodrigues} {et~al.}(2019){Rodrigues}, {Gao}, {Fedynitch}, {Palladino}, \& {Winter}}]{Rodrigues_2019ApJ}
{Rodrigues}, X., {Gao}, S., {Fedynitch}, A., {Palladino}, A., \& {Winter}, W. 2019, \apjl, 874, L29, \dodoi{10.3847/2041-8213/ab1267}

\bibitem[{{Sahakyan}(2018)}]{2018ApJ...866..109S}
{Sahakyan}, N. 2018, \apj, 866, 109, \dodoi{10.3847/1538-4357/aadade}

\bibitem[{{Sikora}(2011)}]{2011IAUS..275...59S}
{Sikora}, M. 2011, in Jets at All Scales, ed. G.~E. {Romero}, R.~A. {Sunyaev}, \& T.~{Belloni}, Vol. 275, 59--67, \dodoi{10.1017/S1743921310015644}

\bibitem[{{Sokolovsky} {et~al.}(2010){Sokolovsky}, {Kovalev}, {Lobanov}, {Savolainen}, {Pushkarev}, \& {Kadler}}]{Sokolovsky_2010arXiv}
{Sokolovsky}, K.~V., {Kovalev}, Y.~Y., {Lobanov}, A.~P., {et~al.} 2010, arXiv e-prints, arXiv:1001.2591, \dodoi{10.48550/arXiv.1001.2591}

\bibitem[{{Stecker} \& {Salamon}(1996)}]{1996SSRv...75..341S}
{Stecker}, F.~W., \& {Salamon}, M.~H. 1996, \ssr, 75, 341, \dodoi{10.1007/BF00195044}

\bibitem[{{Strotjohann} {et~al.}(2019){Strotjohann}, {Kowalski}, \& {Franckowiak}}]{2019A&A...622L...9S}
{Strotjohann}, N.~L., {Kowalski}, M., \& {Franckowiak}, A. 2019, \aap, 622, L9, \dodoi{10.1051/0004-6361/201834750}

\bibitem[{Sánchez-Sáez {et~al.}(2021)Sánchez-Sáez, Reyes, Valenzuela, Förster, Eyheramendy, Elorrieta, Bauer, Cabrera-Vives, Estévez, Catelan, Pignata, Huijse, Cicco, Arévalo, Carrasco-Davis, Abril, Kurtev, Borissova, Arredondo, Castillo-Navarrete, Rodriguez, Ruz-Mieres, Moya, Sabatini-Gacitúa, Sepúlveda-Cobo, \& Camacho-Iñiguez}]{Sanchez-Saez_2021}
Sánchez-Sáez, P., Reyes, I., Valenzuela, C., {et~al.} 2021, The Astronomical Journal, 161, 141, \dodoi{10.3847/1538-3881/abd5c1}

\bibitem[{{Tan} {et~al.}(2023){Tan}, {Liu}, \& {B{\"o}ttcher}}]{Tan2023}
{Tan}, H.-B., {Liu}, R.-Y., \& {B{\"o}ttcher}, M. 2023, arXiv e-prints, arXiv:2311.13873, \dodoi{10.48550/arXiv.2311.13873}

\bibitem[{{Tavecchio} \& {Ghisellini}(2008)}]{2008MNRAS.386..945T}
{Tavecchio}, F., \& {Ghisellini}, G. 2008, \mnras, 386, 945, \dodoi{10.1111/j.1365-2966.2008.13072.x}

\bibitem[{Wang {et~al.}(2020)Wang, Zhang, Sun, \& Liang}]{2020_Zhenjie}
Wang, Z.-J., Zhang, J., Sun, X.-N., \& Liang, E.-W. 2020, The Astrophysical Journal, 893, 41, \dodoi{10.3847/1538-4357/ab7d35}

\bibitem[{{Wang} {et~al.}(2022){Wang}, {Liu}, {Petropoulou}, {Oikonomou}, {Xue}, \& {Wang}}]{Wang_Ze-Rui_2022}
{Wang}, Z.-R., {Liu}, R.-Y., {Petropoulou}, M., {et~al.} 2022, \prd, 105, 023005, \dodoi{10.1103/PhysRevD.105.023005}

\bibitem[{{Willingale} {et~al.}(2013){Willingale}, {Starling}, {Beardmore}, {Tanvir}, \& {O'Brien}}]{Willingale_2013MNRAS}
{Willingale}, R., {Starling}, R.~L.~C., {Beardmore}, A.~P., {Tanvir}, N.~R., \& {O'Brien}, P.~T. 2013, \mnras, 431, 394, \dodoi{10.1093/mnras/stt175}

\bibitem[{{Xue} {et~al.}(2019){Xue}, {Liu}, {Petropoulou}, {Oikonomou}, {Wang}, {Wang}, \& {Wang}}]{Xue_2019}
{Xue}, R., {Liu}, R.-Y., {Petropoulou}, M., {et~al.} 2019, \apj, 886, 23, \dodoi{10.3847/1538-4357/ab4b44}

\bibitem[{{Xue} {et~al.}(2021){Xue}, {Liu}, {Wang}, {Ding}, \& {Wang}}]{Xue_2021}
{Xue}, R., {Liu}, R.-Y., {Wang}, Z.-R., {Ding}, N., \& {Wang}, X.-Y. 2021, \apj, 906, 51, \dodoi{10.3847/1538-4357/abc886}

\end{thebibliography}
\bibliographystyle{aasjournal}


\newpage
\appendix
\renewcommand{\appendixname}{Appendix~\Alph{section}}
Here we present the plots of MCMC fitting under the unified model.
%\section{}

\begin{figure}[ht!]
\plotone{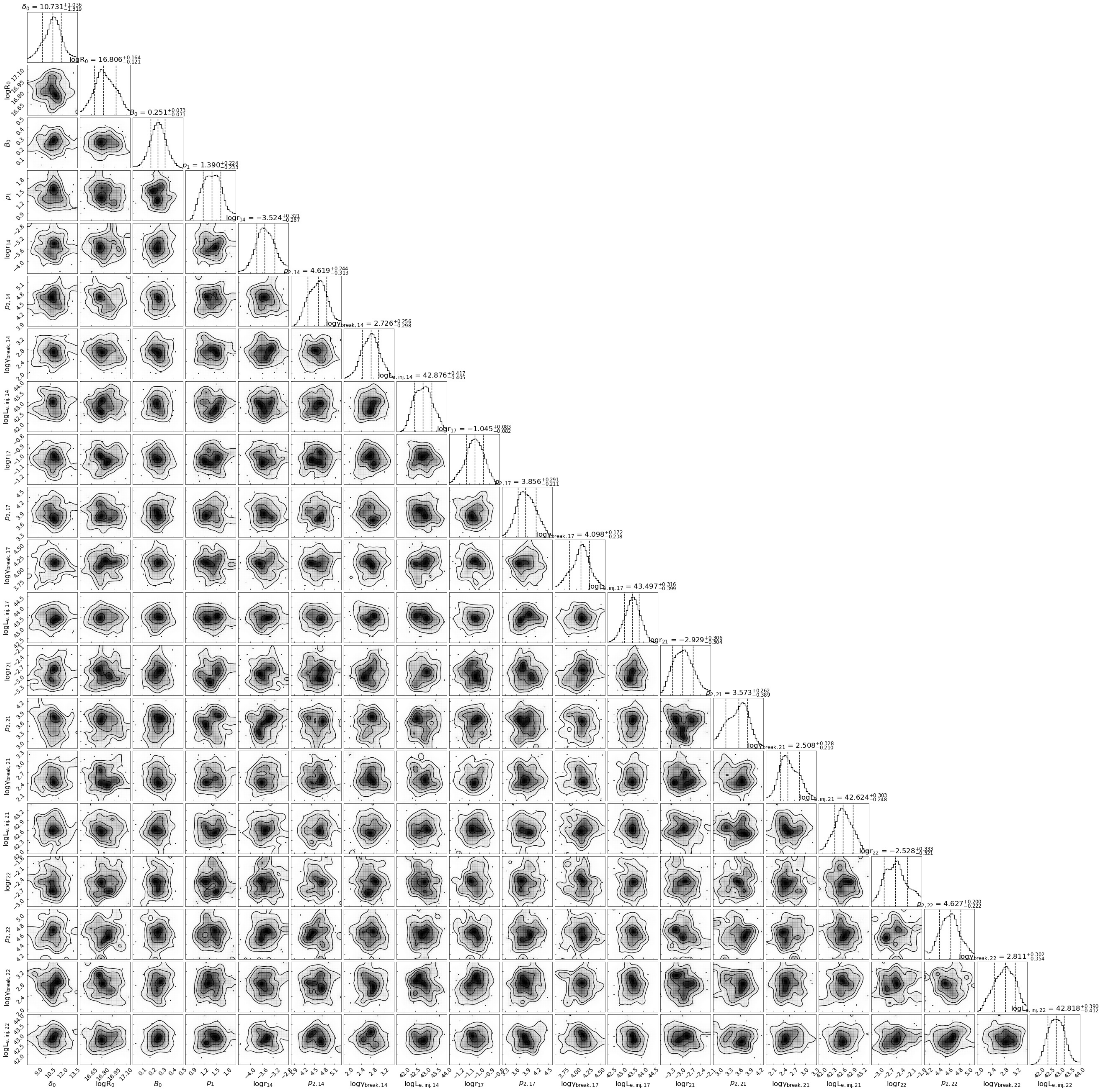}
\caption{The result of MCMC under the unified model\label{fig:general}, the three dotted lines represent the uncertainties based on the 16th, 50th, and 84th percentiles of the samples in the marginalized distributions.}
\end{figure}

\begin{figure}[ht!]
\plotone{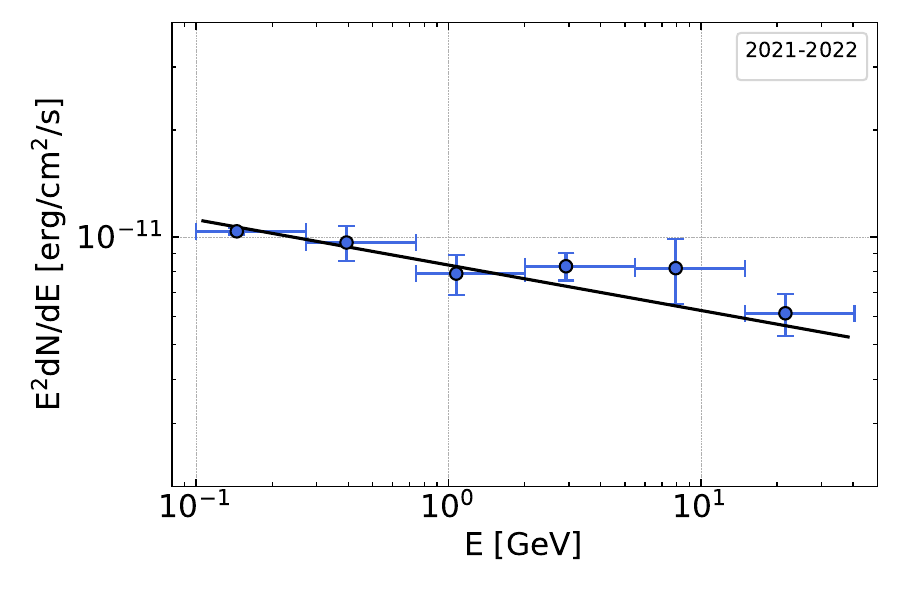}
\caption{The result of SED fitting of Fermi-LAT Observation of TXS 0506+056 during 2021-2022.}
\end{figure}

\begin{figure}[ht!]
\plotone{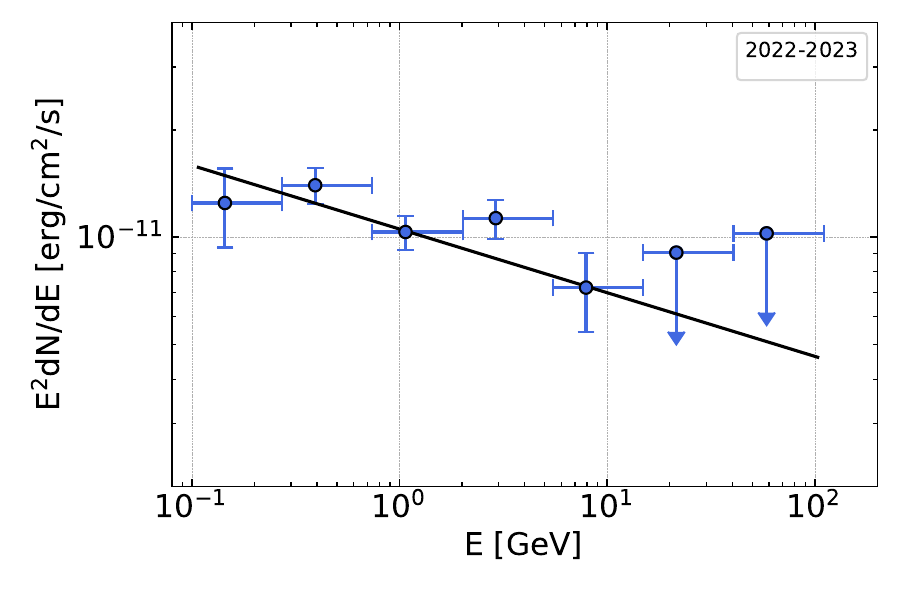}
\caption{The result of SED fitting of Fermi-LAT Observation of TXS 0506+056 during 2022-2023.}
\end{figure}

\begin{figure}[ht!]
\plotone{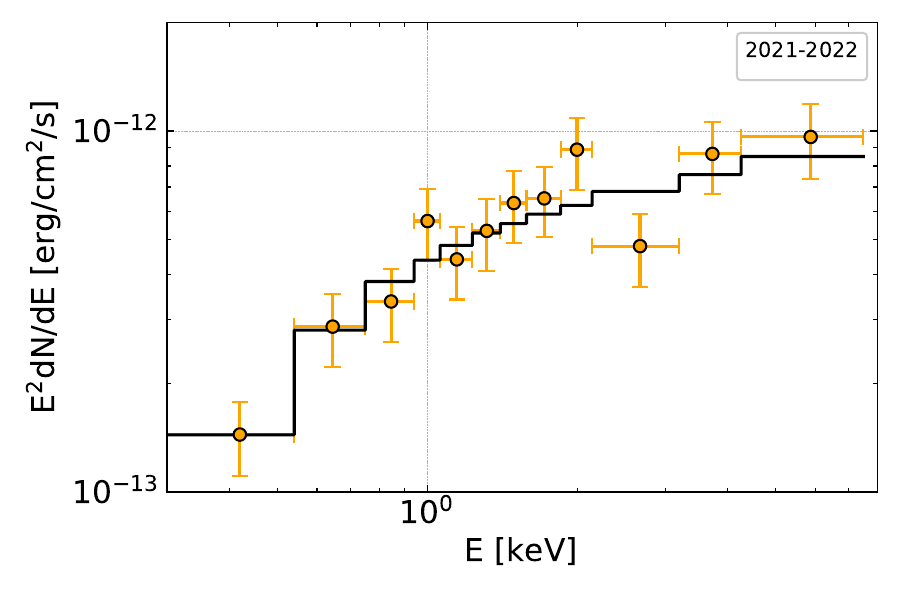}
\caption{The result of SED fitting of Swift-XRT Observation of TXS 0506+056 during 2021-2022.}
\end{figure}

\begin{figure}[ht!]
\plotone{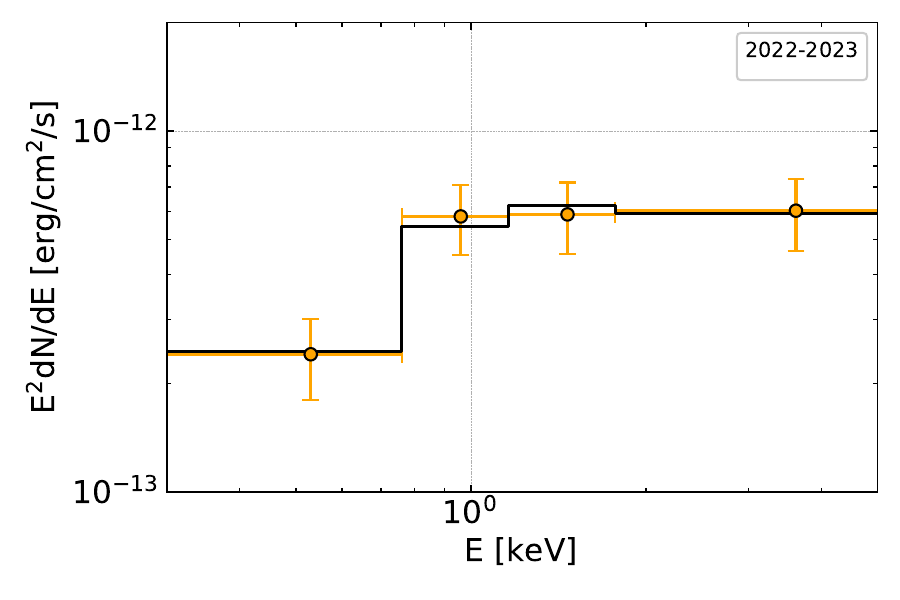}
\caption{The result of SED fitting of Swift-XRT Observation of TXS 0506+056 during 2022-2023.}
\end{figure}

\begin{figure}[ht!]
\plotone{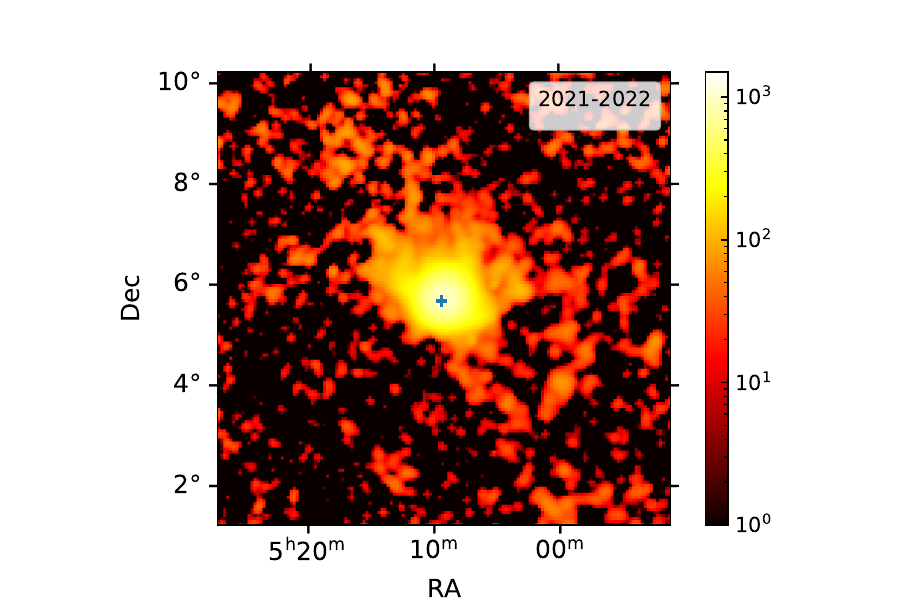}
\caption{The \textit{Fermi} TS map of TXS~0506+056 during 2021-2022.}
\end{figure}

\begin{figure}[ht!]
\plotone{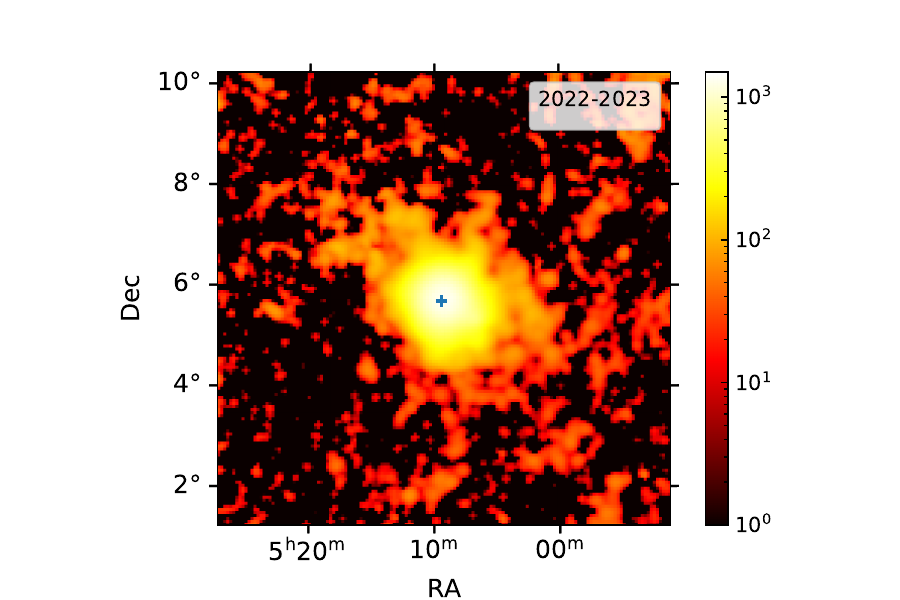}
\caption{The \textit{Fermi} TS map of TXS~0506+056 during 2022-2023.}
\end{figure}

\end{document}